\documentclass[
 reprint,
 amsmath,amssymb,
 aps,
 showkeys,
 superscriptaddress
]{revtex4-2}

\usepackage{graphicx}% Include figure files
\usepackage{dcolumn}% Align table columns on decimal point
\usepackage{bm}% bold math
\usepackage{multirow}
\usepackage{nomencl}
\usepackage{hyperref}% add hypertext capabilities
\usepackage{caption}
\usepackage{subcaption}
\usepackage{gensymb}
\usepackage{xurl}
\usepackage{diagbox}
\usepackage{threeparttable}

\begin{document}

\preprint{APS/123-QED}

\title{Cost and CO$_2$ emissions co-optimisation of green hydrogen production in a grid-connected renewable energy system}

\author{Sleiman Farah}
\altaffiliation{sleiman.farah@mpe.au.dk}
\affiliation{Department of Mechanical and Production Engineering, Aarhus University, Denmark}

\author{Neeraj Bokde}
\affiliation{Department of Mechanical and Production Engineering, Aarhus University, Denmark}
\affiliation{Renewable and Sustainable Energy Research Center, Technology Innovation Institute, Abu Dhabi, 9639, United Arab Emirates}

\author{Gorm Bruun Andresen}
\affiliation{Department of Mechanical and Production Engineering, Aarhus University, Denmark}

% \date{\today}

\begin{abstract}
Green hydrogen is essential for producing renewable fuels that are needed in sectors that are hard to electrify directly. Hydrogen production in a grid-connected hybrid renewable energy plant necessitates smart planning to meet long-term hydrogen trading agreements while minimising costs and emissions.
Previous research analysed economic and environmental impact of hydrogen production based on full foresight of renewable energy availabilty, electricity price, and CO$_2$ intensity in the electricity grid. However, the full foresight assumption is impractical in day-to-day operation, often leading to underestimations of both the cost and CO$_2$ emissions associated with hydrogen production. Therefore, this research introduces a novel long-term planner that uses historical data and short-term forecasts to plan hydrogen production in the day-to-day operation of a grid-connected hybrid renewable energy plant. The long-term planner co-minimises cost and CO$_2$ emissions to determine the hydrogen production for the next day taking into account the remaining hydrogen production and the time remaining until the end of the delivery period, which can be a week, a month, or a year. Extended delivery periods provide operation flexibility, enabling cost and CO$_2$ emissions reductions. Significant reductions in CO$_2$ emissions can be achieved with relatively small increases in the levelised cost. Under day-to-day operation, the levelised cost of hydrogen is marginally higher than that of the full foresight; the CO$_2$ emissions can be up to 60\% higher. Despite a significant portion of the produced hydrogen not meeting the criteria for green hydrogen designation under current rules, CO$_2$ emissions are lower than those from existing alternative hydrogen production methods. These results underscore the importance of balancing cost considerations with environmental impacts in operational decision-making. The results suggest that improvements to the current regulations regarding the green labelling of hydrogen could involve implementing transparent accounting based on hourly CO$_2$ emissions and reducing the specific CO$_2$ emission threshold for green hydrogen production.
\end{abstract}

\maketitle

%\tableofcontents

\section{Introduction}

To fulfill the CO$_2$ reduction commitments of Paris Agreement \cite{parisagreement}, the world, as an international community, needs to search for effective options to control and reduce carbon emissions as well as to maintain a good balance between economic growth and carbon emissions. Most countries are adopting renewable energy policies to reduce CO$_2$ emissions and combat climate change \cite{mangla2020step, razmjoo2020technical}. One of the promising options is to utilise renewable energy to produce green hydrogen that can then replace fossil-fuel, especially in sectors that are difficult to electrify~\cite{HySynergy, orstednews, H2RES}.

In principle, hydrogen is categorised as "green" when produced from renewable energy sources. According to the latest European Union legislation, hydrogen can also be considered as green hydrogen if its production does not generate significant CO$_2$ emissions \cite{Directorate_General_for_Energy}. For example, green hydrogen can be produced using grid electricity when the average renewable energy share exceeds 90\%, the average CO$_2$ emissions intensity of grid electricity is below 18g/MJ, or the hourly electricity price is less than 20~€/MWh \cite{Directorate_General_for_Energy}. Therefore, in countries with significant renewable energy penetration into the electricity grid, green hydrogen production plants are likely to incorporate on-site renewable energy sources, such as wind and solar, as well as being connected to the grid. These plants can optimise their hydrogen production and energy usage by exporting renewable energy to the grid when electricity price is high and importing electricity from the grid when electricity price and CO$_{2}$ intensity are low.

Several green hydrogen projects are under development, especially in northern European countries~\cite{HySynergy, orstednews, H2RES}. Some examples are: the HySynergy project, with an electrolyser capacity exceeding 1.3~GW when the final phase is completed, aims to produce green hydrogen to decarbonise industrial processes in a refinery and for zero-emission mobility~\cite{HySynergy}. The Green Fuels for Denmark project, with an electrolyser capacity exceeding 1.3~GW when the final phase is completed, aims to produce green hydrogen for the production of renewable fuels that can replace fossil-fuels in heavy-duty road transportation, shipping and aviation. The produced hydrogen will be utilised to produce e-fuels such as e-methanol and e-kerosene~\cite{orstednews}. The H2RES project, with an electrolyser capacity of 2~GW, aims to produce green hydrogen to support emission-free road transport in the Greater Copenhagen area and on Zealand island~\cite{H2RES}. These projects utilise electricity from the grid~\cite{HySynergy} as well as renewable energy sources, especially wind power~\cite{orstednews, H2RES}.

The connection to the grid provides flexibility in controlling the power flows in the system, however, the connection to the grid forces these plants to operate according to the electricity market rules. The majority of electricity produced and consumed in Europe is traded in the European power markets. Electricity is traded in several markets on different time scales with the majority of electricity being traded in the day-ahead market. The day-ahead market clears at noon the day before delivery and determines the hourly prices of electricity for the next day. These prices are based on matching bids and offers received from producers and consumers respectively. When participating in the electricity market, both the amount of energy to be either bought or sold during every hour of the upcoming day and the corresponding hourly price (\texteuro/MWh) need to be specified.

\begin{figure}[h]
	\centering{\includegraphics[width=0.45\textwidth] {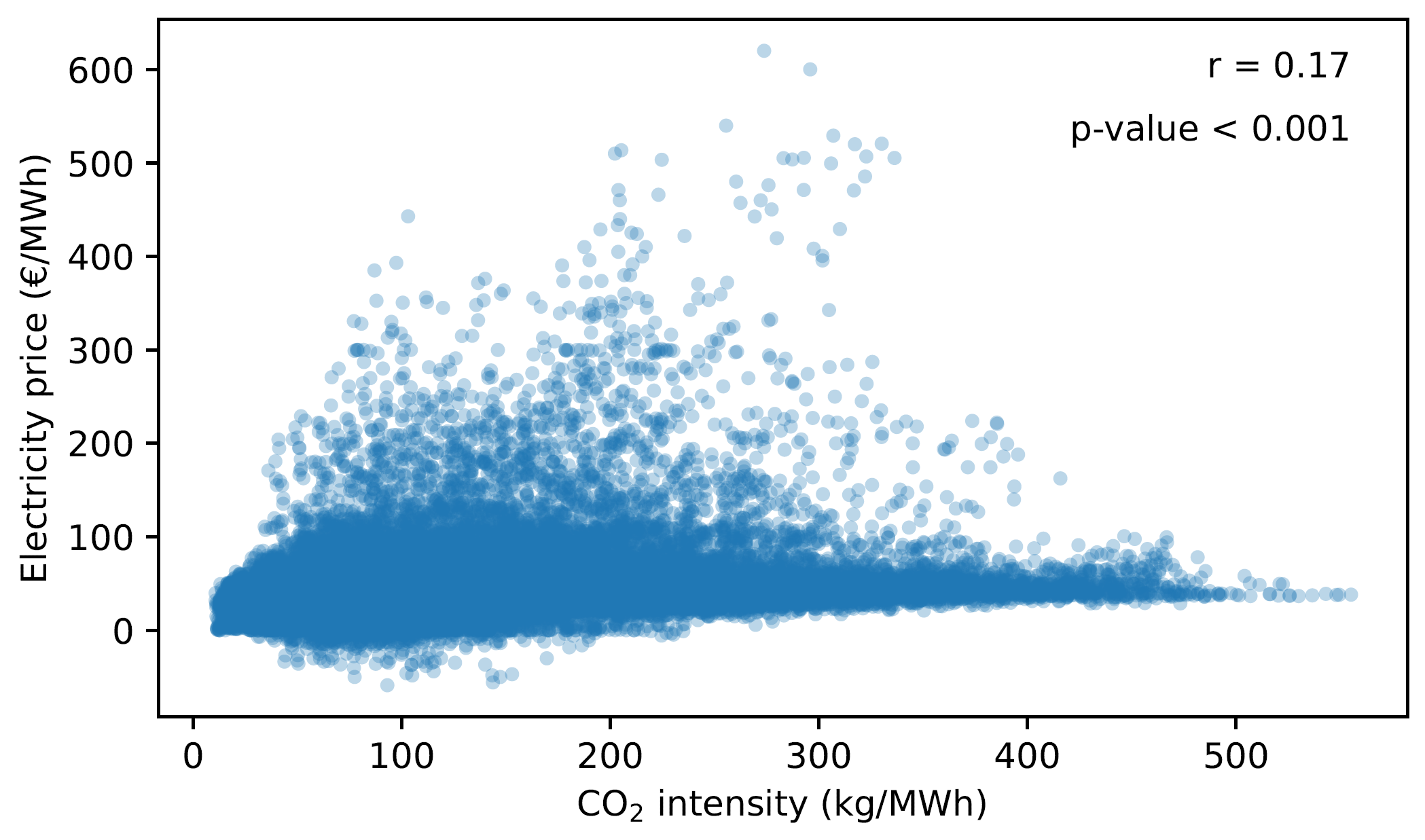}}
	\caption{Correlation of electricity price \cite{entsoe} and CO$_2$eq intensity \cite{energi} in Denmark (bidding zone = DK1, years = 2018--2021).}
	\label{F:electricity_price_vs_co2_intensity}
\end{figure}

Currently, the bidding and operation of hybrid-renewable energy plants are predominantly based on electricity prices in the day-ahead market. This typical bidding and operation approach is suitable for reducing CO$_2$ emissions only when the electricity price and the CO$_2$ intensity are strongly and positively correlated. A strong positive correlation suggests that minimising electricity cost would also reduce CO$_2$ emissions. However, in the case of a strong negative correlation, minimising the electricity cost would increase the CO$_2$ emissions. In this case, hybrid renewable energy plants with hydrogen production need to co-minimise the electricity cost and the CO$_2$ emissions to satisfy the requirements for producing green hydrogen. Co-optimisation is also necessary when the electricity price and the CO$_2$ intensity are weakly correlated. For instance in Denmark, the weak correlation is expressed by a low correlation coefficient ($r=0.17$) and by a wide range of CO$_2$ intensity for a given electricity price (Fig.~\ref{F:electricity_price_vs_co2_intensity}). The weak correlation suggests that minimising the electricity cost alone does not necessarily achieve a reduction in CO$_2$ emissions.

The day-ahead electricity market provides a platform for electricity to be traded daily, however, such a market does not exist for hydrogen which is typically traded through hydrogen purchase agreements. In these agreements, a specified mass of hydrogen should be produced by the end of a specified period, such as a day, a week, a month, or a year. The required production of hydrogen over an extended period provides additional operation flexibility of the renewable hybrid plant. However, the extended period raises the challenge of devising a long-term plan for hydrogen production across the extended delivery period. The challenge is due to unreliable forecast of hourly electricity price, CO$_2$ intensity, and renewable energy availability for long forecast horizons. Consequently, a long-term plan for hydrogen production entails determining a suitable daily mass of hydrogen to be produced throughout the entire delivery period. In other words, the total hydrogen mass that needs to be produced during a delivery period is predetermined, however, the mass that should be produced at different days is unknown. A simple long-term plan involves equal daily production of hydrogen; the same mass of hydrogen is produced every day, and the co-optimisation determines the most suitable hours for hydrogen production within the day. Despite the selection of the most suitable hours every day, such a plan fails to align with the variable electricity price, CO$_2$ intensity, and renewable energy production in the long-term. For instance, on days with high electricity prices, exporting electricity to the grid is favoured over producing hydrogen, which can be produced on a later day, still within the same delivery period, when the electricity price and CO$_2$ intensity are low. In this case, the plan of equal daily production of hydrogen fails to benefit from delaying the production even though the delay does not compromise fulfilling the hydrogen purchase agreement.

A limited number of studies have analysed both cost and CO$_2$ emissions associated with hydrogen production~\cite{ENGSTAM2023117458, SORRENTI2023113033}. A system consisting of an electrolyser connected to a wind farm and supported by the electricity grid was analysed considering three operating strategies~\cite{ENGSTAM2023117458}. Two strategies minimised cost and CO$_2$ emissions, whereas a third strategy minimised the electricity cost considering electricity from the grid only. The system was supply-driven as different electrolyser full load hours (FLH) were considered without specific demand profiles for hydrogen and electricity. A trade-off between cost and CO$_2$ emissions reduction was identified. Grid electricity supporting the production of hydrogen from wind energy could reduce the levelised cost of hydrogen. Increasing the wind to electrolyser capacity ratio could reduce both the cost and CO$_2$ emissions~\cite{ENGSTAM2023117458}. However, the system optimisation is based on full foresight for an entire year which overestimates knowledge about future wind energy, electricity prices and CO$_2$ intensities, and does not capture the day-to-day operation where knowledge about future values is limited to a short horizon. Similarly, four scenarios for grid-connected hydrogen production were considered in the Danish industrial park GreenLab skive~\cite{SORRENTI2023113033}. The system operation considered detailed models of various components and a predefined hydrogen demand profile. The system operation prioritised supplying renewable electricity to the electrolyser to satisfy the hydrogen demand. Surplus renewable electricity was exported to the grid whereas electricity was imported from the grid when renewable electricity was insufficient to satisfy the hydrogen demand. The scenarios considered different ratios of renewable capacity and electrolyser capacity to identify the most suitable system that minimises cost and CO$_2$ emissions~\cite{SORRENTI2023113033}. However, the predefined hydrogen demand profile seems to reduce the potential of minimising cost and CO$_2$ emissions, especially when the profile does not align favourably with the availability of renewable energy, low electricity price and CO$_2$ intensity.

While these studies analyse cost and CO$_2$ emissions associated with hydrogen production, they also reveal important shortcomings; they ignore realistic day-to-day system operation and assume a predefined hydrogen demand profile. However, some aspects of these shortcomings can be overcome using the graphical method for co-minimisation of electricity cost and CO$_2$ emission which was developed for power-to-X storage applications~\cite{bokde2020graphical}. This method considers day-to-day operation without being restricted to a predefined hydrogen demand profile. The method consisted of three steps: The first involved generating a scatter plot of the latest hourly electricity prices and CO$_2$ intensities for a designated delivery period, such as a day, a week, a month, or a year. The second involved plotting a line at a specific angle to balance the trade-off between electricity price and CO$_2$ intensity. The line was then shifted until the desired number of points, which represent the required hours of operation, is within the region defined by the main axes of the plot and the inclined line. The third involved generating a scatter plot of the forecast of hourly electricity prices and CO$_2$ intensities for the next day. The points that fell beneath the line produced in the second step were selected for importing electricity from the grid to charge the power-to-X storage. The method required correcting the selected hours by adding hours that were above the sloped line when the required hours of operation could not be met by the hours below the line. This correction added more hours of operation towards the end of the delivery period~\cite{bokde2020graphical}. The graphical method is simple and easy to implement; however, by ignoring the temporal relationship between the selected hours for importing electricity, this method is unable to consider operational constraints, such as ramp-up to full capacity, that prohibit limiting the hours to the ones beneath the sloped line. The method also required operation correction that increases the operation at the end of the delivery period. This correction can adversely affect the economical and environmental performances when high electricity prices and CO$_2$ emissions occur towards the end of a delivery period.

Consequently, to enhance existing literature, this research presents a grid-connected hybrid renewable energy system for hydrogen production and electricity trading. The system simulation considers day-to-day operation to co-optimise electricity cost and CO$_2$ emissions. The daily hydrogen production profile varies as determined by the co-optimisation. The novelty is a long-term planning method that calculates the hydrogen mass that should be produced the next day without relying on long-term forecasts. The aim from simulating this system is to quantify the cost and CO$_2$ emissions trade-off for hydrogen production considering realistic day-to-day operation with short-term forecast and technical operation constraints of the hydrogen production plant, including ramp-up limits of the electrolyser. The aim is also to analyse the impact of production flexibility provided by longer delivery periods on the cost and CO$_2$ emissions.
 
The subsequent sections of the paper are structured as follows: Section~\ref{S:Methodology} presents the hydrogen production system and the main features of the optimisation model. Section~\ref{S:Inputs} presents the main inputs to the optimisation model. Section~\ref{S:Results} presents the main results and discussion on the system performance. Finally, Section~\ref{S:Conclusions} presents a summary of the research and concluding remarks.

\section{Methodology}
\label{S:Methodology}

\subsection{System description}

\begin{figure}
\centering{\includegraphics[width=0.45\textwidth]{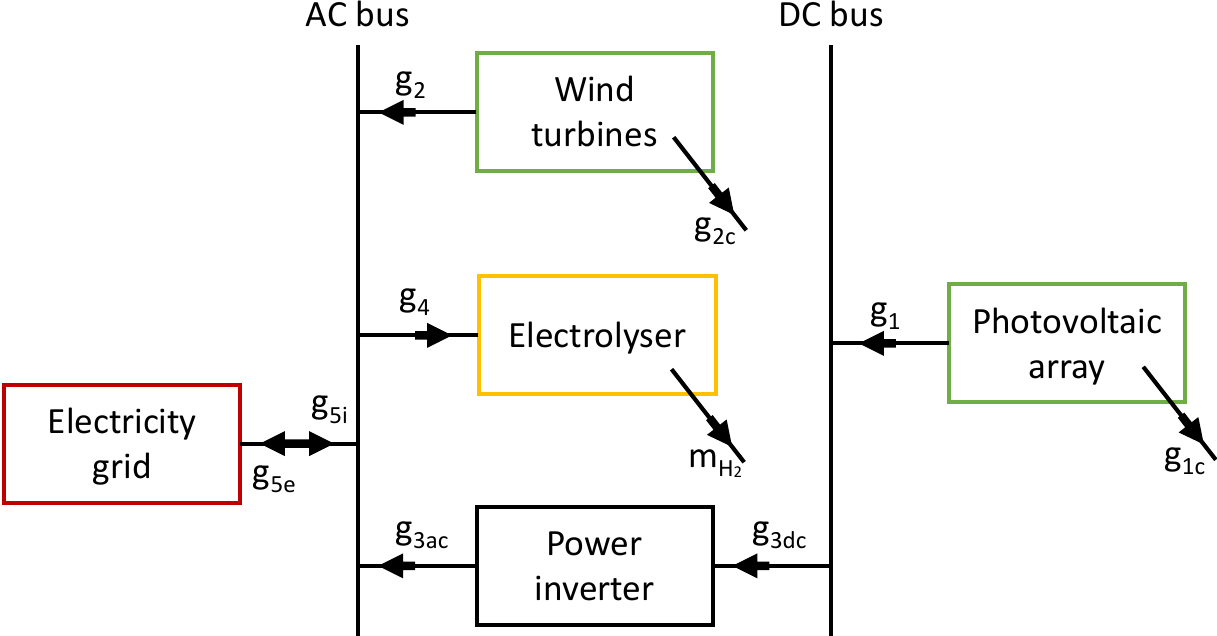}}
	\caption{Schematic of hydrogen production plant.}
	\label{F:Schematic_of_green_hydrogen_production system}
\end{figure}

The hydrogen production plant consists of a solar photovoltaic array and wind turbines as two power sources, an electrolyser for hydrogen production, and an inverter, as shown in Fig.~\ref{F:Schematic_of_green_hydrogen_production system}. The solar power source is connected to the direct current (DC) bus, whereas the wind power source and the electrolyser are connected to the alternating current (AC) bus. Electric power can flow through the inverter from the DC bus to the AC bus. The hydrogen production plant can export and import electricity from the grid which is connected to the AC bus.

\begin{figure*}
	\centering{\includegraphics[width=0.95\textwidth]{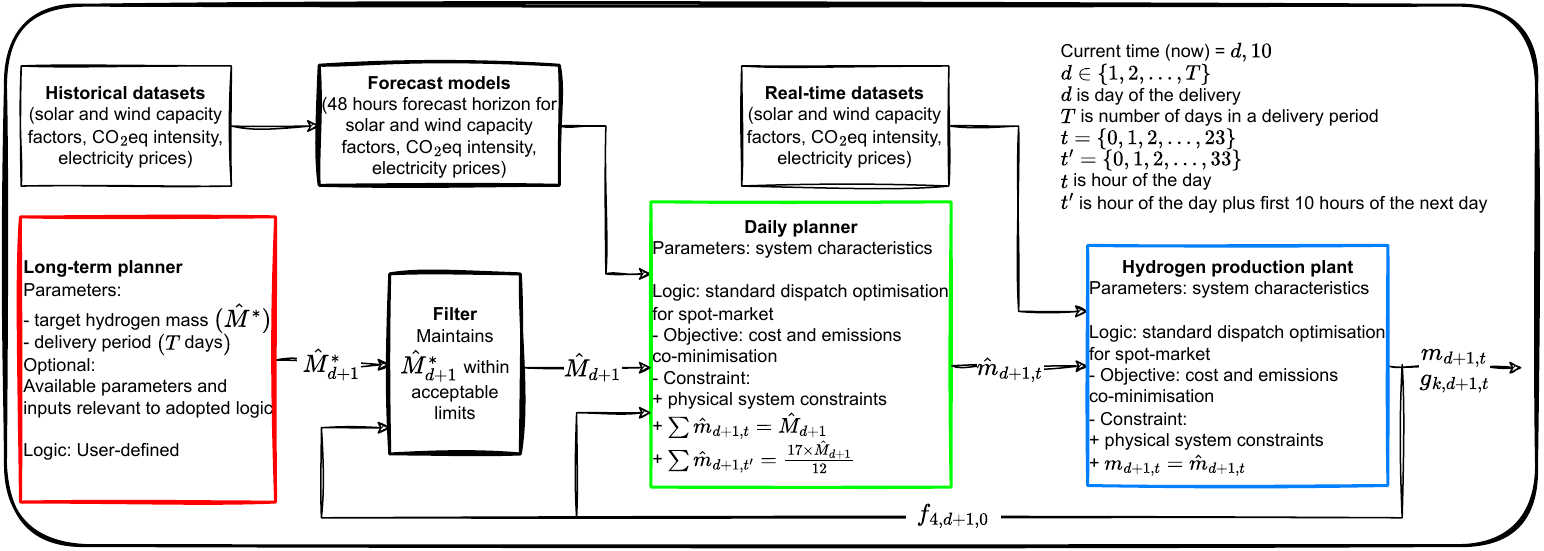}}
	\caption{System operation logic.}
	\label{F:system operation logic}
\end{figure*}

The system operates for an entire year with the aim of repeatedly producing a specified hydrogen mass within defined delivery periods which can be a day, a week, a month, or a year. To produce the required mass of hydrogen within a delivery period, the hydrogen production plant is supported by two control modules, the long-term and the daily planners (Fig.~\ref{F:system operation logic}). Within a delivery period, the time of hydrogen production is unrestricted, as long as the specified mass of hydrogen is produced by the end of the delivery period.

The long-term planner determines the overall hydrogen mass to be produced in the upcoming day, while the daily planner creates a hydrogen production plan for each hour of the upcoming day. The output of the daily planner also defines the plan of hourly power imports and exports from the grid for the upcoming day. Typically, this plan is submitted to the grid operator for consideration in the dispatch planning for the next day. However, considering that forecasts of wind and solar capacity factors do not match exactly the actual capacity factors, maintaining the hydrogen production plan as well as the power imports and exports to the grid cannot always be satisfied simultaneously. For instance, based on forecast that significant renewable energy will be available, a hydrogen production plan may consist of producing hydrogen and exporting electricity to the grid both at full capacity. However, the actual available renewable power might be lower than the forecast values and the system would only be able to satisfy one or the other of the power flows; i.e., either maintain full hydrogen production without meeting the requirement of exporting electricity to the grid at full capacity, or maintain exporting electricity to the grid at full capacity and ignore the hydrogen production. In this research, maintaining the planned production of hydrogen is considered the highest priority for the business model, and the planned power exports and imports from the grid are assumed to be compensated for in the intra-day market for the same electricity prices of the day-ahead market.

Every day at 10:00~am, the long-term planner provides the required hydrogen mass ($\hat M^*_{d+1}$) which passes through a filter that ensures the required hydrogen mass is within the acceptable limits of the hydrogen production plant ($\hat M^*_{d+1} \xrightarrow{}\hat M_{d+1}$). These limits are defined by the minimum and maximum mass of hydrogen that the plant can produce in a day considering the ramping-up and ramping-down characteristics as well as the capacity factor ($f_{4,d+1,0}$) of the electrolyser at the beginning of the upcoming day. The electrolyser capacity factor at the end of the present day ($f_{4,d+1,0}$) is known despite being a future value as the operation of the electrolyser for the rest of the present day is predetermined by the hydrogen production plan from the previous day.

The filtered mass $\hat M_{d+1}$ is increased considering 10~additional hours for hydrogen production at the beginning of the day after the upcoming day. This addition ensures that the daily planner does not optimise the system operation strictly for upcoming day, completely ignoring that the system still has to operate favourably the day after the upcoming day. The daily planner allocates the increased hydrogen production optimally for each hour of upcoming day and the 10 hours in the day after the upcoming day with the constraint that total hydrogen production in the upcoming day is equal to $\hat M_{d+1}$. The inclusion of 10 additional hours influence the state of operation of the electrolyser; for instance, the addition ensures that the electrolyser is not shut down at the end of the upcoming day if optimal operation for the first 10 hours of the day after the upcoming day favours the electrolyser being operational. The importance of this addition is more pronounced if energy storage were included in the system. Without considering additional hours from the day after the upcoming day the system would tend to utilise all the energy available in the energy storage; leaving no energy to be used the day after the upcoming day when a better usage of that energy might be possible.

The daily planner optimisation considers a model of the hydrogen production plant and inputs from forecast models for solar and wind capacities, as well as forecasts for electricity prices and CO$_2$eq intensities. For simplicity, the forecast models are assumed perfect which means that forecast and real values are identical. Additionally, the daily planner optimisation takes into account the capacity factor ($f_{4,d+1,0}$) of the electrolyser at the start of the upcoming day. The daily planner optimises hydrogen production and power flows with the objective of co-minimising cost and CO$_2$ emissions as shown in the objective function in Eq.~\ref{E:objective_function}.
The hydrogen production plan created by the daily planner is executed by the hydrogen production plant that considers real solar and wind capacity factors as well as electricity price and CO$_2$ intensity. In the general framework, the real data does not match exactly the forecast data, however, considering the simplifying assumption of perfect forecast, short-term forecast and real data are identical.

The main optimisation features of the hydrogen production plant, the daily planner and the long-term planner are presented in Section~\ref{S:System model}.

\subsection{\label{S:System model}System model}
\subsubsection{Hydrogen production plant model}
The objective function of the co-optimisation is a weighted sum of both electricity and operation costs and CO$_2$ emissions cost as shown in Eq.~\ref{E:objective_function}
\begin{equation}
\min_{g_{k,t}}
    \hat {C}_{\alpha} = \alpha\hat{C}_{CO_2} + (1-\alpha)(\hat{C}_{e} + \hat{C}_o)
\label{E:objective_function}
\end{equation}
where $\alpha$ is the CO$_2$ weighting factor ($0\le\alpha\le 1$). The variables $\hat{C}_{CO_2}$, $\hat{C}_{e}$ and $\hat{C}_o$ are the cost of CO$_2$ emissions, the cost of importing electricity from the grid, and the operation cost of various components in the system respectively. For $\alpha = 0$, the optimisation results correspond to the lowest operation cost, whereas for $\alpha = 1$, the optimisation results correspond to hydrogen production that generates the least  CO$_2$ emissions. For $0<\alpha < 1$, the optimisation co-minimises cost and CO$_2$ emissions. The optimisation model of hydrogen production plant considers actual solar and wind capacity factors, CO$_2$ intensity and electricity prices as inputs. The main operational requirement is to follow the hydrogen production plan optimised by the daily planner, as shown in Eq.~\ref{E:plant_constraint_1}
\begin{equation}
    \label{E:plant_constraint_1}
    m_{d+1, t} = \hat{m}_{d+1, t}
\end{equation}
where $m_{d+1, t}$ and $\hat{m}_{d+1, t}$ are the actual produced hydrogen and the hydrogen mass that should be produced at every hour respectively. The details of cost calculation and physical constraints of the plant are presented in Appendix~\ref{A:Optimisation model details}.
\subsubsection{Daily planner model}
The daily planner shares the same objective function (Eq.~\ref{E:objective_function}) and model details (Appendix~\ref{A:Optimisation model details}) as the hydrogen production plant. However instead of actual values, the daily planner considers short-term forecast of solar and wind capacity factors, CO$_2$ intensity and electricity prices as inputs. The optimisation period considered by the daily planner is 34~hours starting from the upcoming day until 10:00~am the day after the upcoming day; the first 14~hours of forecasts which correspond for the remaining of today are ignored.  The main operational requirement is that the total mass of hydrogen that should be produced the upcoming day is equal to the filtered value ($\hat{M}_{d+1}$) specified by the long-term planner, as shown in Eq.~\ref{E:daily_planner_constraint_1}
\begin{equation}
    \label{E:daily_planner_constraint_1}
    \sum \hat{m}_{d+1,t}= \hat{M}_{d+1}
\end{equation}
\subsubsection{Long-term planner model}
\begin{figure*}
	\centering{\includegraphics[width=0.95\textwidth]{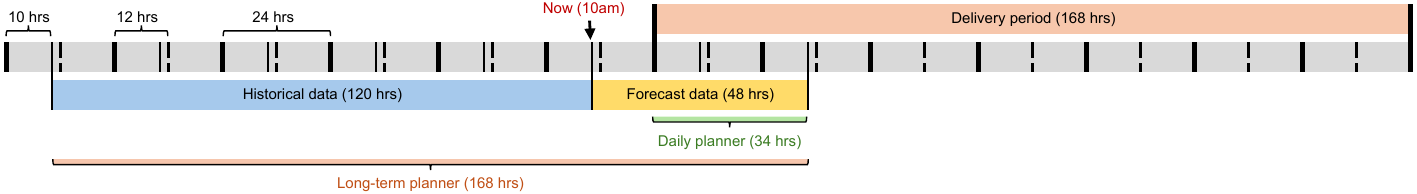}}
	\caption{Illustration of data and time periods utilised by the long-term and daily planners (example of first day of a weekly delivery).}
	\label{F:long_term_planner_approach}
\end{figure*}

The long-term planner shares the same objective function (Eq.~\ref{E:objective_function}) and model details (Appendix~\ref{A:Optimisation model details}) as the hydrogen production plant and the daily planner. However, the novel approach of the long-term planner is that the optimisation considers both short-term forecasts and the most recent actual solar and wind capacity factors, CO$_2$ intensity and electricity prices, as shown in Fig.~\ref{F:long_term_planner_approach}. A typical approach would require long-term input forecasts, which are inherently unreliable, to produce the outputs, however, the proposed approach replaces long-term forecasts by encompassing historical inputs with short-term forecast. Optimisation based on the short-term forecast and historical inputs is statistically equivalent to optimisation based on long-term forecasts. This hypothesis is justified as forecast models are constructed based on historical data to predict future values, i.e., statistical characteristics of past and future data are identical.\\
An important feature of the long-term planner is the consideration of a variable time window that shrinks progressively every day. The shrinking window is adaptive to the remaining of both hydrogen target and time to the end of the delivery period. The shrinking window ensures that the hydrogen target is achieved by the end of the delivery period without any additional production corrections.
The hydrogen production planning for a new delivery period starts on the day prior to the start of the delivery period, where the long-term planner considers a time window encompassing both forecasts and historical data that is identical to the temporal extent of the delivery period. The main operational requirement is that the total mass of hydrogen that should be produced in this time window is equal to the target hydrogen mass ($\hat{M}^*$) that should be produced within the delivery period. The optimisation provides the optimal distribution of hydrogen production, which includes the hydrogen mass that should be produced the upcoming day ($\hat{M}^*_{d+1}$).
In the subsequent days of a delivery period, the temporal extent of historical data is reduced by one day every day so that the time window encompassing both forecasts and historical data is identical to the remaining time of the delivery period. The operational requirement is updated so that the total mass of hydrogen that should be produced in the reduced time window is equal to the remaining of the target hydrogen mass. The optimisation provides the optimal distribution of hydrogen production, which includes the hydrogen mass that should be produced the upcoming day ($\hat{M}^*_{d+1}$), similar to the process on the day prior to the start of the delivery period. This process is repeated until the third to last day of a delivery period, i.e., there are two days left to plan for, where the reduced time window does not include the historical data anymore and the inputs consists of the short-term forecasts only. On the second to last day of a delivery period, the long-term planner provides the remaining of the target hydrogen mass. On the last day of a delivery period, the long-term planner starts a new planning cycle for the upcoming delivery period.

\section{Inputs}
\label{S:Inputs}
The system performance is analysed for various deliveries, namely, daily, weekly, monthly, and yearly; the corresponding hydrogen mass required is presented in Table~\ref{T:adopted_hydrogen_demand}.
\begin{table}
\caption{\label{T:adopted_hydrogen_demand}Hydrogen demand for various delivery periods.}
\begin{threeparttable}
\begin{tabular}{l c c}
    \hline
    Delivery periods  & Hydrogen mass\tnote{\textdagger} (kg) \\
    \hline
    Day & 296 ($\times 365$)\\
    Week & 2071 ($\times 52$)\\
    Month & 8877 ($\times 12$)\\
    Year & 108000 ($\times 1$)\\
    \hline
\end{tabular}
\begin{tablenotes}
    \item[\textdagger] Based on 6000 full load hours of the electrolyser within a year.
\end{tablenotes}
\end{threeparttable}
\end{table}
The system capacities and electrolyser characteristics are presented in Table~\ref{T:system_capacities} and~\ref{T:electrolyser_parameters} respectively. These values are not optimised and they are selected to showcase the differences between benchmark and day-to-day performances. An inverter efficiency of 90\% is assumed to account for part load operation.
\begin{table}
\caption{\label{T:system_capacities}System capacities.}
\begin{threeparttable}
\begin{tabular}{l c}
    \hline
    Description & Capacity (MW)\\
    \hline
    Photovoltaic & 1.0\\
    Wind & 1.0\\
    Electrolyser & 1.0\\
    Inverter & 1.0\\
    Grid connection & 1.0\\
    \hline
\end{tabular}
% \begin{tablenotes}
%     \item[\textdagger] my notes here.
% \end{tablenotes}
\end{threeparttable}
\end{table}
\begin{table}
\caption{\label{T:electrolyser_parameters}Electrolyser parameters \cite{DanishEnergyAgency_inverter}.}
\begin{threeparttable}
\begin{tabular}{l c c}
    \hline
    Description & Value & Unit\\
    \hline
    Efficiency ($\eta_{LHV_{H_2}}$) & 60\tnote{\textdagger} & \%\\
    \\[-0.8em]
    Ramp rate up ($rru$) & 100.0 & \%/h\\
    Ramp rate up cold ($rruc$) & 50.0 & \%/h\\
    Ramp rate down ($rrd$) & 100.0 & \%/h\\
    \\[-0.8em]
    Capacity cost ($c_{e}$) & 700 & $10^3\times$\texteuro/MW\\
    \\[-0.8em]
    Lifetime ($T_{e}$) & 10 & years\\
    \hline
\end{tabular}
\begin{tablenotes}
    \item[\textdagger] Based on low heat value ($LHV_{H_2}=120$~MJ/kg). Refer to Appendix~\ref{A:electrolyser_efficiency} for more details. 
\end{tablenotes}
\end{threeparttable}
\end{table}

\begin{figure}
    \centering
    \begin{subfigure}[b]{0.45\textwidth}
        \includegraphics[height=5cm]{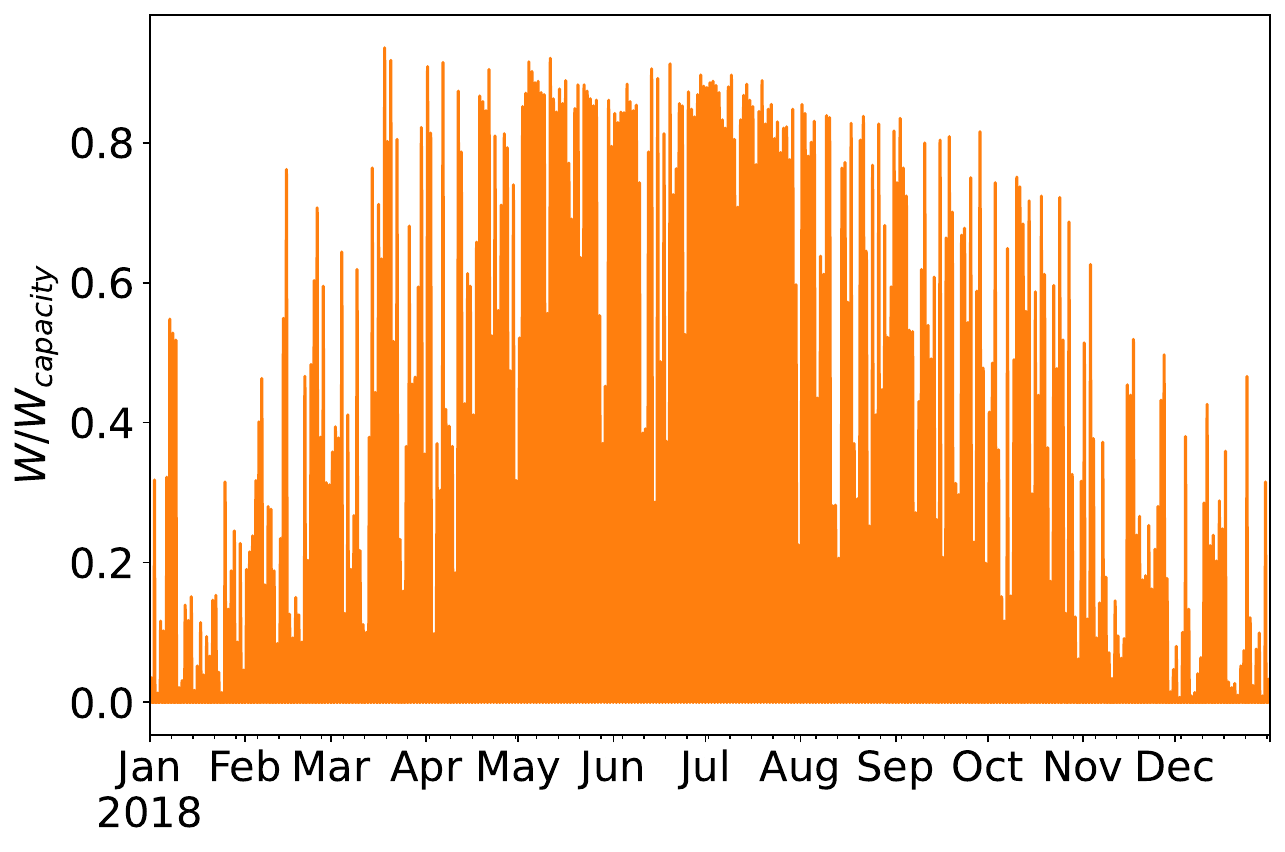} % Adjust height as needed
        \caption{Solar capacity factor}
        \label{Fig:solar_capacity_factors}
    \end{subfigure}
    \quad
    \begin{subfigure}[b]{0.45\textwidth}
        \includegraphics[height=5cm]{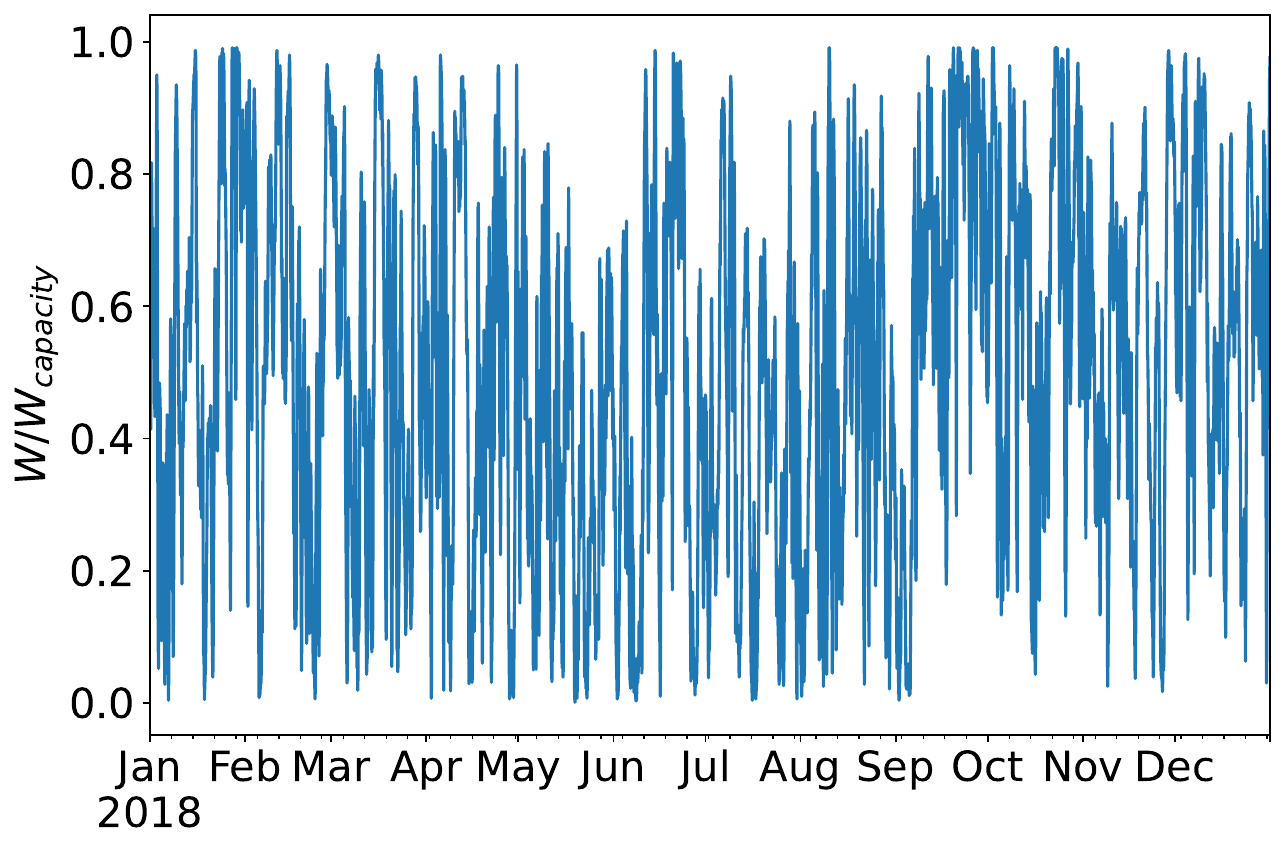} % Adjust height as needed
        \caption{Wind capacity factor}
        \label{Fig:wind_capacity_factors}
    \end{subfigure}
    
    \begin{subfigure}[b]{0.49\textwidth}
        \includegraphics[height=5cm]{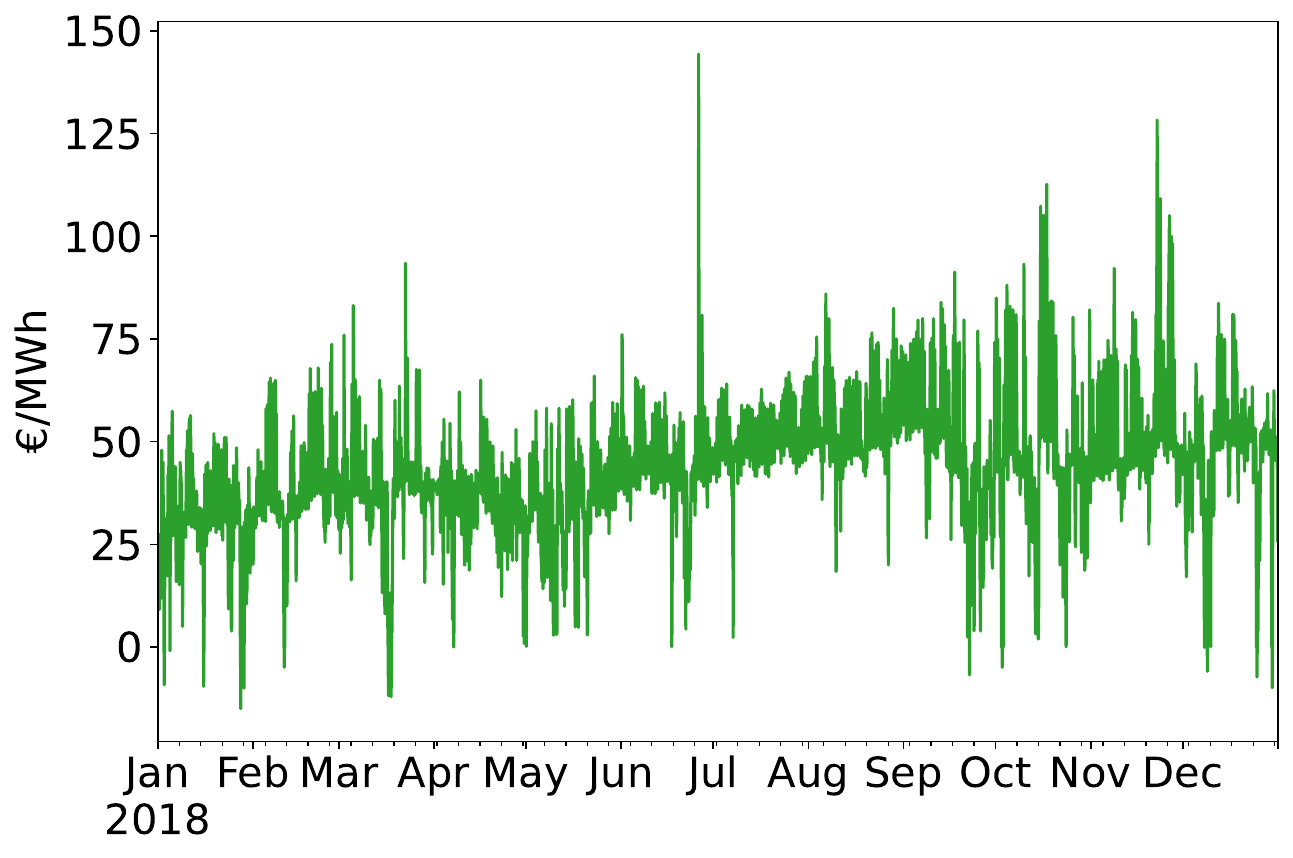} % Adjust height as needed
        \caption{Electricity prices}
        \label{Fig:electricity_prices}
    \end{subfigure}
    \quad
    \begin{subfigure}[b]{0.49\textwidth}
        \includegraphics[height=5cm]{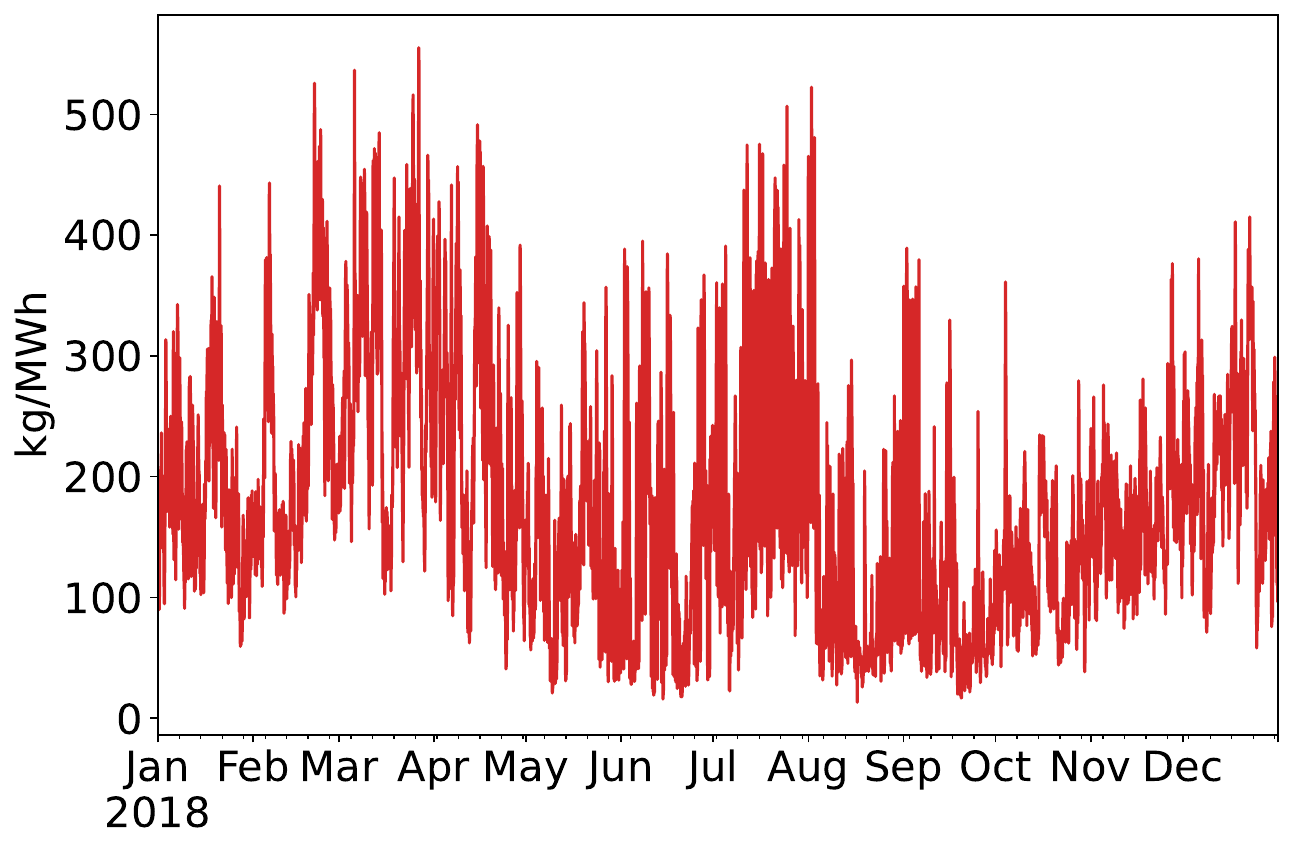} % Adjust height as needed
        \caption{CO$_2$ intensity}
        \label{Fig:co2_equivalent_intensity}
    \end{subfigure}
    \caption{Time series for the year 2018, north of Jutland, Denmark, latitude: 56.6135\degree, longitude: 8.9328\degree).}
    \label{F:Time series for the year 2018}
\end{figure}

The time series of solar and wind capacity factors, electricity price, and CO$_2$ intensity are presented in Fig.~\ref{F:Time series for the year 2018}. 

\section{Results and discussion}
\label{S:Results} 
\subsection{System performance}
The system performance considering day-to-day operation is compared to benchmark performance which is calculated by optimising the energy flows considering full foresight spanning the entire year. The benchmark performance is also based on the objective of co-optimising the cost and CO$_2$ emissions (Eq.~\ref{E:objective_function}). The comparison considers the levelised cost of hydrogen and specific CO$_2$ emissions for different delivery periods. The levelised cost of hydrogen considers capital and electricity costs as well as the lost revenue that could have been obtained if the system were operated for trading electricity only. The specific CO$_2$ emissions are calculated as the ratio of total CO$_2$ emissions to total hydrogen production in one year. The results in this section are for the year 2018, while the results corresponding to 2019--2021 are presented in Appendix~\ref{A:additional_results}.

\begin{figure}
  \begin{subfigure}[t]{0.49\textwidth}
    \includegraphics[width=\textwidth]{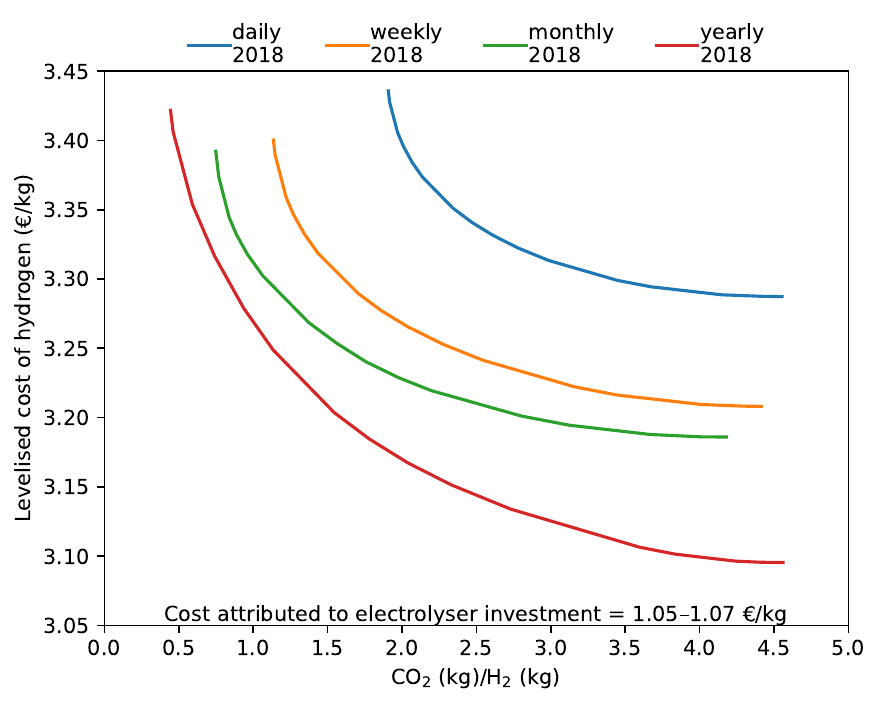}
    \caption{Benchmark}
    \label{F:benchmark_lcoh_co2}
  \end{subfigure}
  \hfill
  \begin{subfigure}[t]{0.49\textwidth}
    \includegraphics[width=\textwidth]{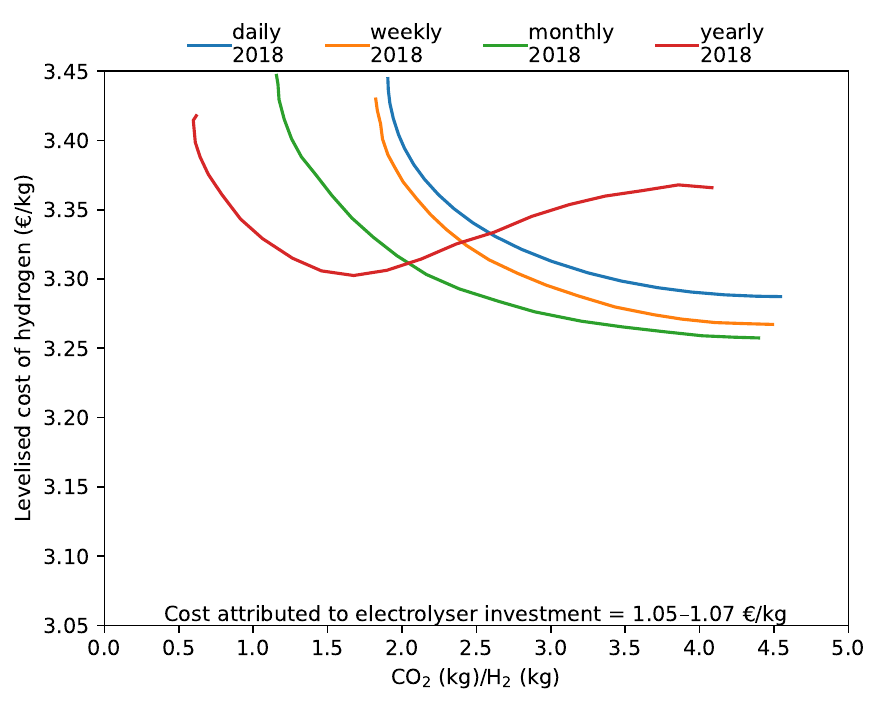}
    \caption{Day-to-day}
    \label{F:day-to-day_lcoh_co2}
  \end{subfigure}
  \caption{Impact of specific CO$_2$ emissions and delivery frequencies on the levelised cost of hydrogen.}
  \label{F:impact_levelised_cost_h2}
\end{figure}

The benchmark results show a general tendency for the levelised cost to decrease as the specific CO$_2$ emission increases (Fig.~\ref{F:benchmark_lcoh_co2}). The trade-off between the specific CO$_2$ emission and the levelised cost is driven by the change of the CO$_2$ weighting factor ($\alpha$) in the objective function (Eq.~\ref{E:objective_function}). A large reduction of specific CO$_2$ emissions can be achieved with a relatively small increase of the levelised cost (Table~\ref{T:Ranges of specific CO$_2$ emissions and levelised cost}). As expected, longer delivery periods are beneficial for system performance, with the yearly delivery having the lowest specific emission and cost. 

The day-to-day results also show a general tendency for the levelised cost to decrease as the specific CO$_2$ emission increases for all delivery frequencies except the yearly delivery (Fig.~\ref{F:day-to-day_lcoh_co2}). For daily, weekly, and monthly deliveries, the system achieves the intended performance, a trade-off between the levelised cost and specific CO$_2$ emissions. For the yearly delivery, the system achieves a similar trade-off for low specific CO$_2$ emissions, however, the system fails to continue reducing the cost as the specific CO$_2$ intensity increases. This result means that when the CO$_2$ weighting factor is reduced with the aim of achieving lower costs, the total cost increases instead of decreasing. This result is due to the inability of foreseeing electricity prices, CO$_2$ intensity, and renewable energy availability far into the future. Consequently, despite the optimisation achieving local minimum costs every day the total cost over the entire year increases. Further analysis that explains the day-to-day performance for the yearly delivery is presented in Appendix~\ref{A:analysis_of_day_to_day_performance_for_yearly_delivery}.
\begin{table}
    \centering
    \caption{Ranges of specific CO$_2$ emission and levelised cost of hydrogen}
    \begin{tabular}{l c c}
        \hline
        Delivery & Specific CO$_2$ emissions & Levelised cost\\
        period & range (kg CO$_2$/kg H$_2$) & range (€/kg H$_2$)\\
        \hline
        \multicolumn{3}{c}{Benchmark}\\
        \hline
        Day & 2.65 & 0.16\\
        Week & 3.28	& 0.20\\
        Month & 3.40 & 0.21\\
        Year & 4.12	& 0.34\\
        \hline
        \multicolumn{3}{c}{Day-to-day}\\
        \hline
        Day & 2.64 & 0.16\\
        Week & 2.67 & 0.16\\
        Month & 3.24 & 0.19\\
        Year & 3.49 & 0.12\\
        \hline
    \end{tabular}
    \label{T:Ranges of specific CO$_2$ emissions and levelised cost}
\end{table}

\begin{figure}
    \includegraphics[width=0.49\textwidth]{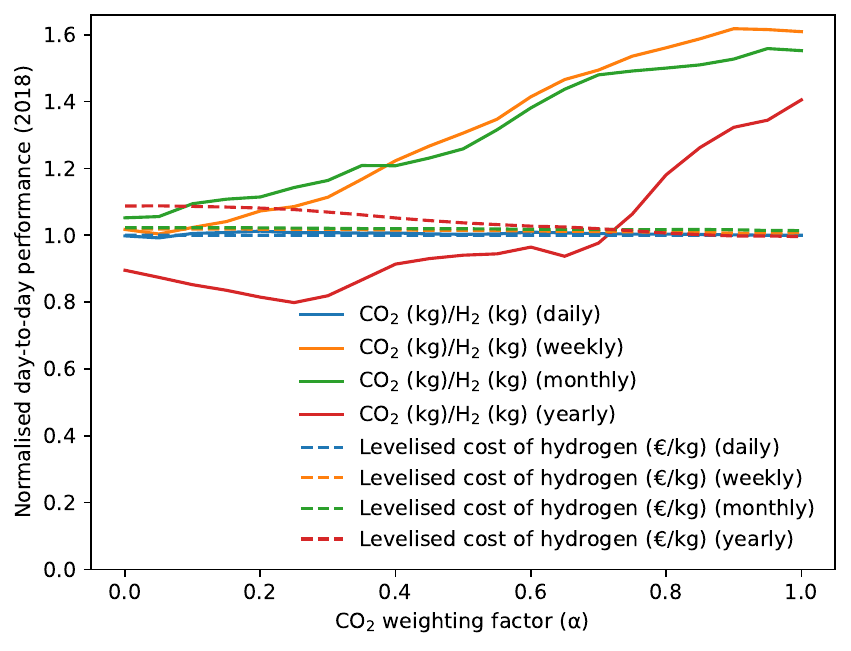}
    \caption{Comparison of benchmark and day-to-day performances for different deliveries. Normalised values are calculated as the ration of the day-to-day performance to the corresponding benchmark performance.}
    \label{F:comparison_of_benchmark_and_day_to_day_performances_for_different_deliveries}
\end{figure}

The comparison of benchmark and day-to-day performances is based on the normalised values presented in Fig.~\ref{F:comparison_of_benchmark_and_day_to_day_performances_for_different_deliveries}. The normalised values are calculated as the ratio of day-to-day performances for specific CO$_2$ emission and levelised cost of hydrogen to the corresponding benchmark performances for the same delivery periods and CO$_2$ weighting factors. The value one indicates that benchmark and day-to-day performances are identical. For low CO$_2$ weighting factors where the co-optimisation aims to achieve lower costs, the day-to-day performance for the levelised costs of hydrogen are slightly higher than one. For high CO$_2$ weighting factors where the co-optimisation aims to achieve lower CO$_2$ emissions, the emissions based on the day-to-day performance can exceed by more than 20\% up to 60\% the benchmark performance. These results indicate that assessment of hydrogen production based on full foresight underestimates slightly the levelised cost of hydrogen; nevertheless, the assessment can significantly underestimates the environmental impact. This underestimation can have a significant adverse impact for planning for the green transition and achieving the CO$_2$ emission reduction targets.
\subsection{Green hydrogen evaluation}
This section analysis the system performance from the perspective of green hydrogen production considering the latest regulation in the European Union~\cite{Directorate_General_for_Energy}. For brevity and focusing on the CO$_2$ emissions associated with hydrogen production inline with~\cite{Directorate_General_for_Energy}, results of yearly delivery and CO$_2$ weighting factor equal one that achieve the lowest CO$_2$ emissions are considered. In accordance with the guidelines outlined in~\cite{Directorate_General_for_Energy}, a set of rules depending on the operational context dictate whether the electricity utilised to produce hydrogen can be renewable energy, hence, the produced hydrogen is green hydrogen. Consequently, the system performance from the perspective of green hydrogen production is analysed considering the rules that are the most relevant to the presented system. Two particular rules that allow considering electricity from the grid as renewable energy at all times depend on the annual average of renewable energy penetration ($>$ 90\%) or the annual average of CO$_2$ intensity ($<$ 64.8~kg/MWh). In 2023 in Denmark, the highest energy penetration from wind and solar and the lowest CO$_2$ intensity are 69.4\% and 185~kg/MWh respectively~\cite{nowtricity}. In the bidding zone DK1, where more renewable energy is available, the renewable energy penetration and the lowest CO$_2$ intensity are 74\% and 91~kg/MWh respectively (data source~\cite{declaration_grid_mix}).

\begin{figure}
  \begin{subfigure}[t]{0.49\textwidth}
    \includegraphics[width=\textwidth]{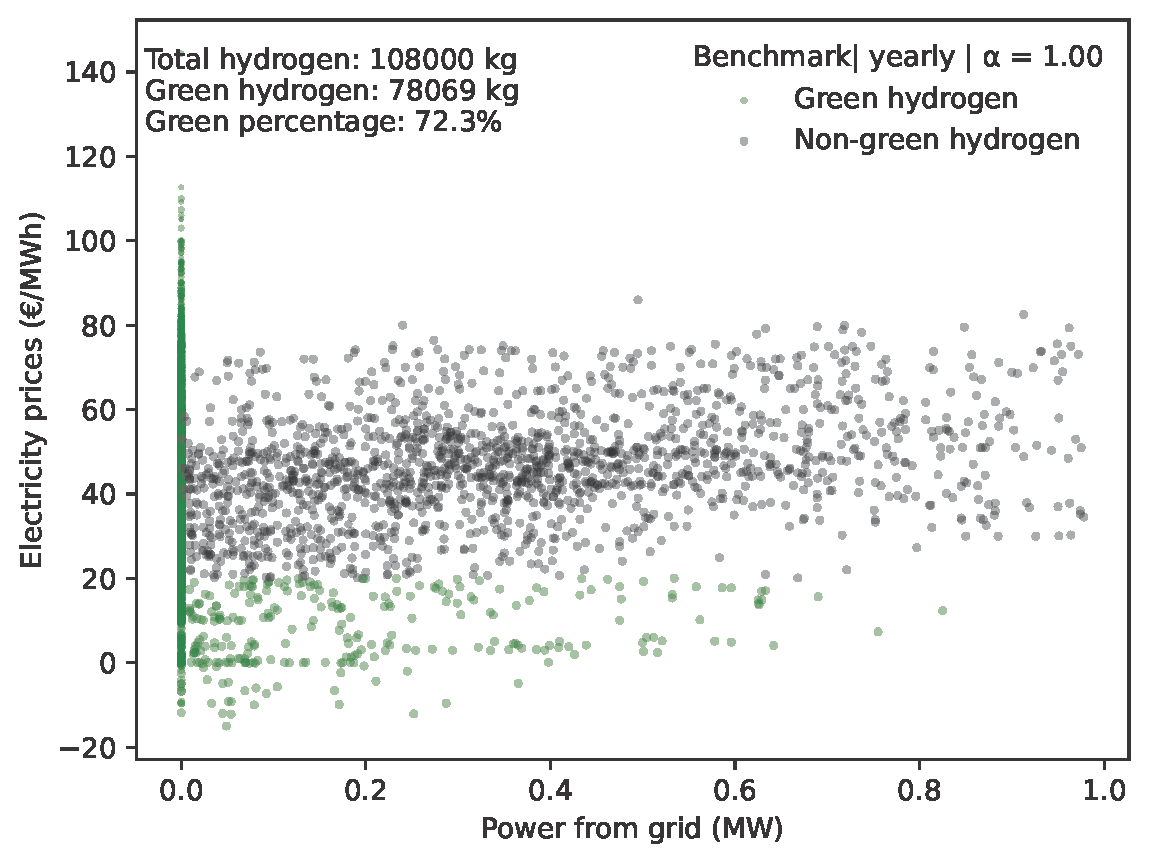}
    \caption{Benchmark}    \label{F:h2_production_vs_electricity_from_grid_and_electricity_price_y2018_yearly_co2w_1.0_benchmark}
  \end{subfigure}
  \hfill
  \begin{subfigure}[t]{0.49\textwidth}
    \includegraphics[width=\textwidth]{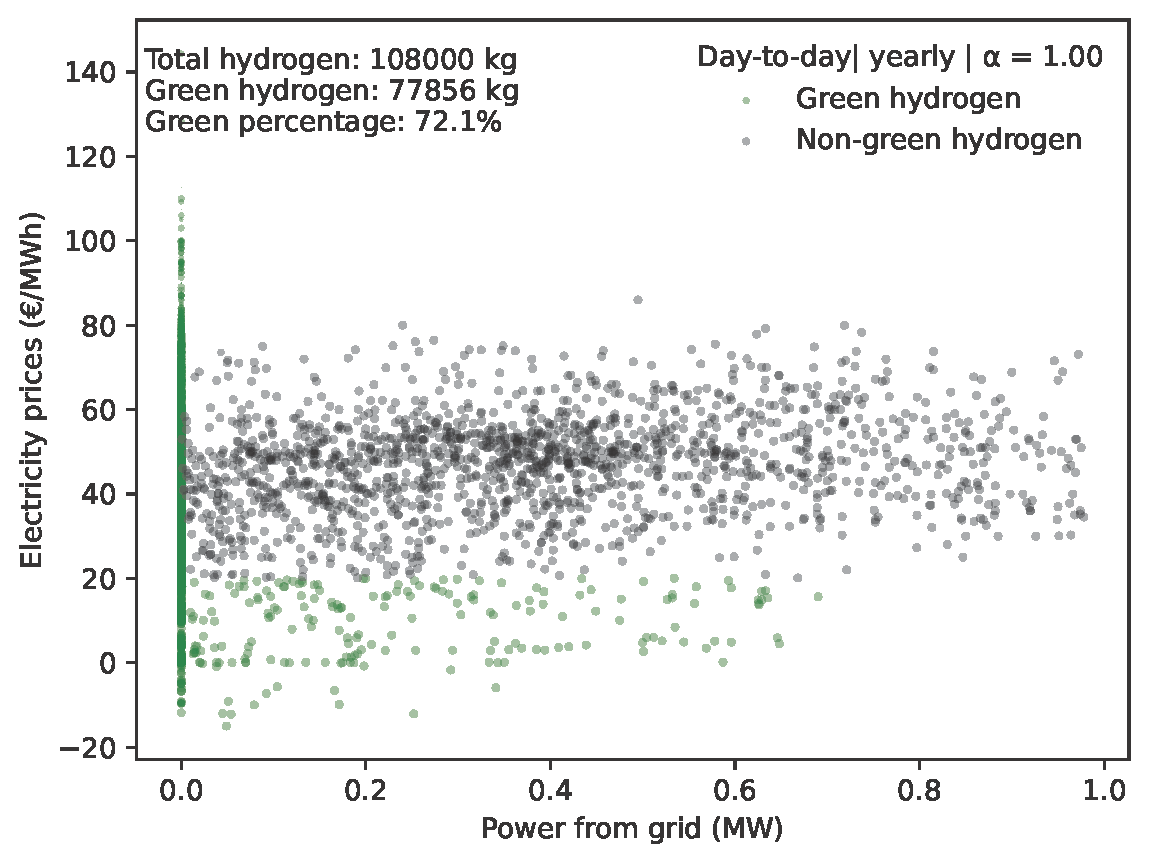}
    \caption{Day-to-day}    \label{F:h2_production_vs_electricity_from_grid_and_electricity_price_y2018_yearly_co2w_1.0_shrinking_window}
  \end{subfigure}
  \caption{Hydrogen production distribution considering electricity prices and power from the grid.}
\label{F:hydrogen_production_distribution_considering_electricity_prices_and_power_from_the_grid}
\end{figure}

In principle, these values indicate that electricity from the grid cannot be considered as renewable energy for the purpose of green hydrogen production. Consequently, green hydrogen from on-site renewable energy considering benchmark and day-to-day operation is 67.9\% and 68.8\% of the total hydrogen produced respectively. However, regardless of the annual average of renewable energy penetration and CO$_2$ intensity, electricity from the grid for the purpose of hydrogen production is considered renewable energy when the electricity price is less than 20~€/MWh~\cite{Directorate_General_for_Energy} (Fig.~\ref{F:hydrogen_production_distribution_considering_electricity_prices_and_power_from_the_grid}). Hence, considering both on-site renewable energy and electricity from the grid the green hydrogen share increases slightly to 72.3\% and 72.1\%.
The remainder of produced hydrogen is considered non-renewable hydrogen according to the rules set in \cite{Directorate_General_for_Energy}. However, the CO$_2$ emissions generated per kilogram of non-renewable hydrogen are 1.5~kg and 2.1~kg for benchmark and day-to-day performance respectively. These values are lower than the the CO$_2$ generated from hydrogen produced by coal gasification (14--15~ kg), steam methane reforming (8--10~kg), or biomass gasification (2--3~kg)~\cite{GH2_Facts}. Therefore, despite that part of the produced hydrogen is not considered as green hydrogen, all the produced hydrogen has significantly lower CO$_2$ emissions than existing alternative hydrogen production methods.

\begin{figure}
  \begin{subfigure}[t]{0.45\textwidth}
    \includegraphics[width=\textwidth]{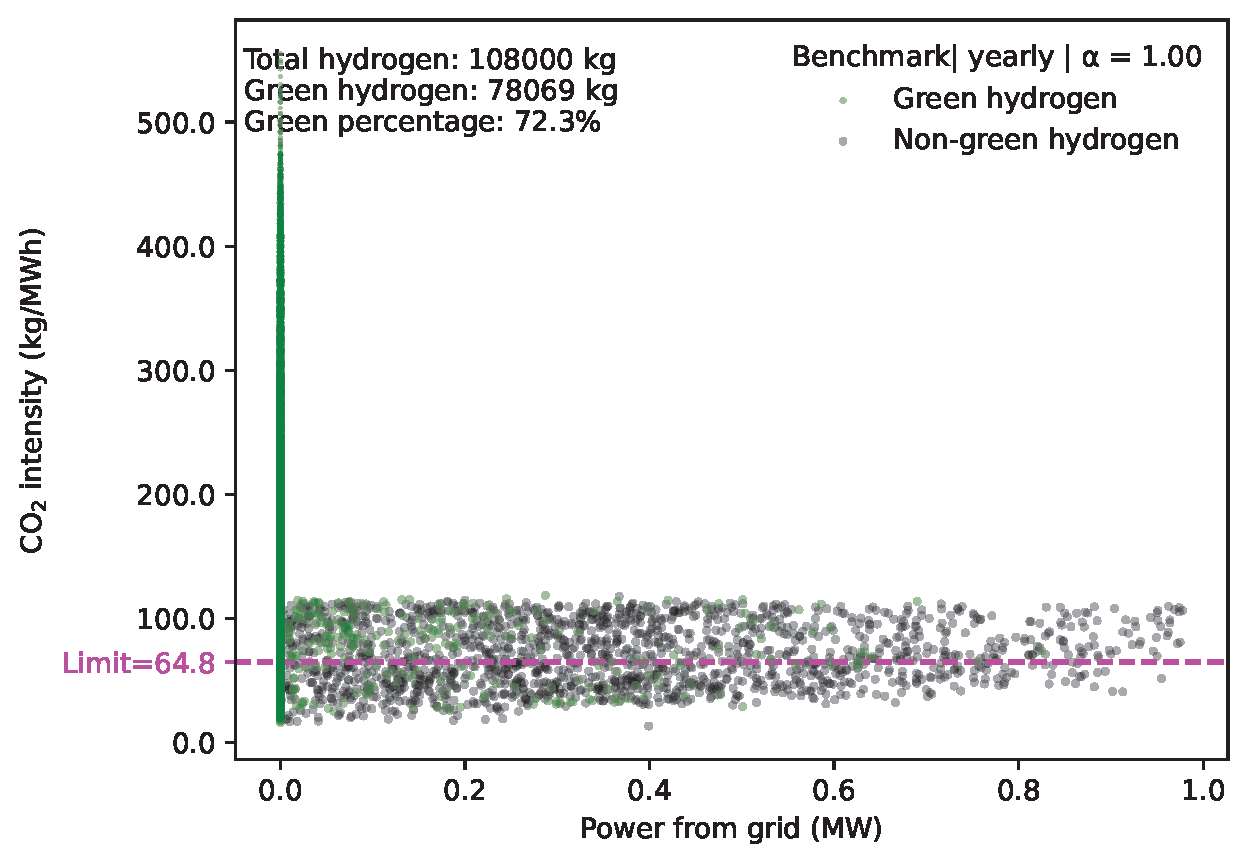}
    \caption{Benchmark}
    \label{F:h2_production_vs_electricity_from_grid_and_co2_intensity_y2018_yearly_co2w_1.0_benchmark}
  \end{subfigure}
  \hfill
  \begin{subfigure}[t]{0.45\textwidth}
    \includegraphics[width=\textwidth]{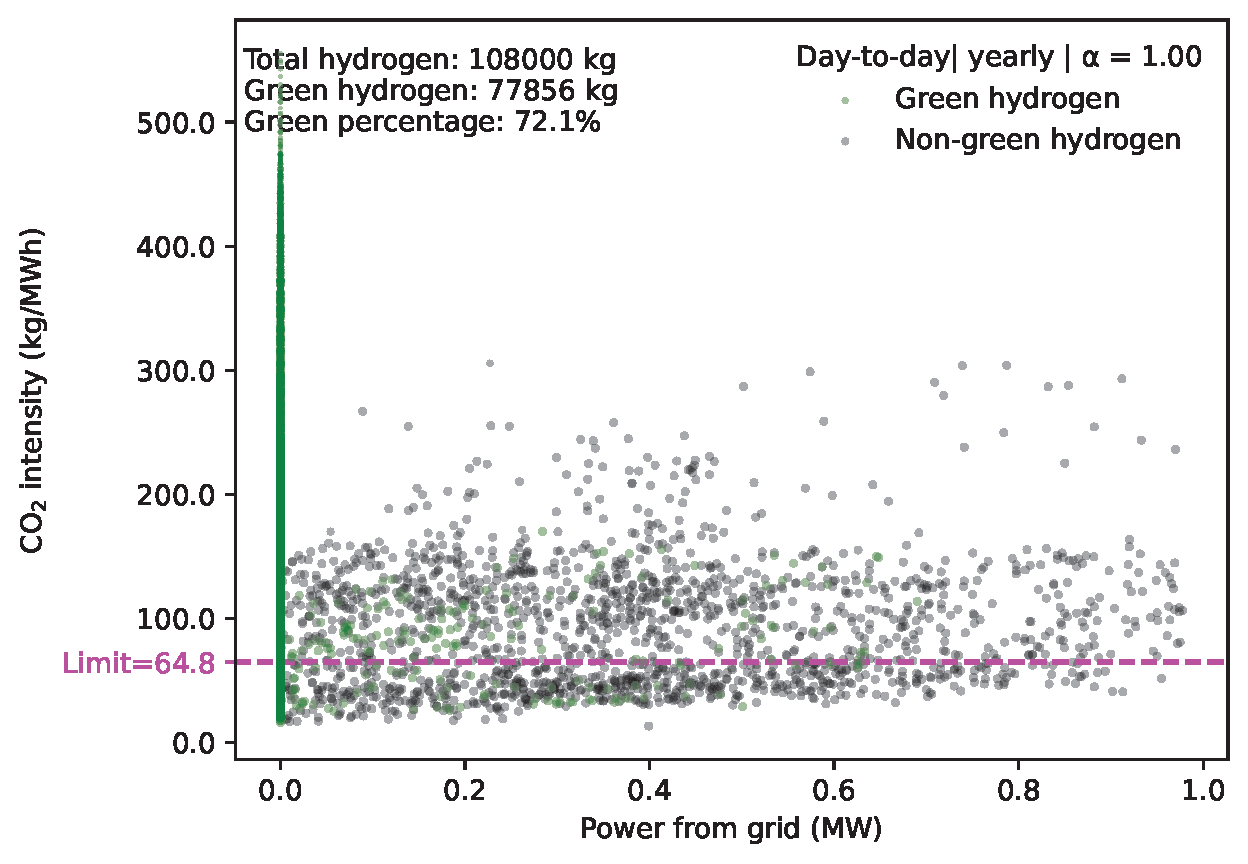}
    \caption{Day-to-day}    \label{F:h2_production_vs_electricity_from_grid_and_co2_intensity_y2018_yearly_co2w_1.0_shrinking_window}
  \end{subfigure}
  \caption{Hydrogen production distribution considering CO$_2$ intensity and power from the grid.}
\label{F:hydrogen_production_distribution_considering_co2_intensity_and_power_from_the_grid}
\end{figure}

Analysing the rules set in~\cite{Directorate_General_for_Energy} for green hydrogen production reveal that they could have been more transparent. The rule of annual average of renewable energy penetration being greater than 90\% implies that up to 10\% of the utilised energy can be from non-renewable sources. Hence, one could argue for claiming an additional 10\% of green hydrogen, particularly in situations where CO$_2$ emissions from peak-power plants, such as gas power plants, are low. Similarly, the annual average CO$_2$ intensity being less than 64.8~kg/MWh implies that generating CO$_2$ emissions in the production of green hydrogen is permissible; for a 60\% electrolyser efficiency, this value corresponds to specific emissions average equal to 3.6~kg of CO$_2$ per kg of hydrogen produced. Consequently, recalling the results in Fig.~\ref{F:impact_levelised_cost_h2}, one could argue for claiming all the produced hydrogen as green hydrogen, especially when the co-optimisation objective is CO$_2$ emissions minimisation. The specific emissions is approximately 0.5~kg of CO$_2$ per kilogram of hydrogen produced. Despite this specific emissions value being seven times lower than the value permitted by the 64.8~kg/MWh rule, the current regulation does not allow claiming all the produced hydrogen as green. One could also argue that a more transparent and stricter regulation should consider hydrogen produced when hourly CO$_2$ intensity is below the 64.8~kg/MWh limit as green hydrogen. Considering hourly CO$_2$ intensity means that a significant share of the currently considered non-green hydrogen should be considered as green hydrogen; all the non-green hydrogen (black dots in Fig.~\ref{F:hydrogen_production_distribution_considering_co2_intensity_and_power_from_the_grid}) produced below the 64.8~kg/MWh limit should also be considered as green hydrogen. 
Further, the rule considering a limit on electricity price instead CO$_2$ intensity is peculiar. The results in Fig.~\ref{F:hydrogen_production_distribution_considering_co2_intensity_and_power_from_the_grid} show that to some extent both green hydrogen and non-green hydrogen are produced at the same CO$_2$ intensity level. This observation indicates that having a limit on the electricity price does not achieve the intended purpose, which is producing green-hydrogen only when the CO$_2$ is low. This outcome is inline with the observations that electricity prices and CO$_2$ emissions are weakly correlated (recall Fig.~\ref{F:electricity_price_vs_co2_intensity}).
These results indicate that while the rules set in \cite{Directorate_General_for_Energy} help developing the green hydrogen industry, some improvements can be suggested:
\begin{itemize}
    \item Adoption of transparent accounting based on hourly CO$_2$ emissions associated with hydrogen production regardless of the annual average levels of renewable energy penetration and CO$_2$ intensity in the electricity grid
    \item Adoption of a lower specific CO$_2$ intensity as the current limit of 3.6~kg of CO$_2$ per kg of hydrogen produced is comparable to producing hydrogen from biomass gasification. A recommended upper limit for the specific CO$_2$ intensity is 1.5~kg of CO$_2$ per kg of hydrogen produced. Considering the results from different years (Figs.~\ref{F:impact_levelised_cost_h2}, \ref{F:Impact of specific CO$_2$ emissions and delivery frequency on the levelised cost H$_2$ (year 2019)}--\ref{F:Impact of specific CO$_2$ emissions and delivery frequency on the levelised cost H$_2$ (year 2021)}), this limit is achievable for all long delivery periods when the objective is to minimise the CO$_2$ emissions without the need for excessively large solar and wind capacities. This limit can be reduced gradually as renewable energy penetration into the electricity grid increases.
\end{itemize}
These recommendations increase the economical benefits for green hydrogen producers with reduced impact on the environment compared to the current rules. 

\section{Summary and conclusions}
\label{S:Conclusions}
Addressing climate change challenges requires reducing CO$_2$ emissions from all sectors. Several projects are being developed to produce green hydrogen, which is necessary for the production of renewable fuels that can be utilised in sectors that are difficult to decarbonise. According to the rules in the European Union, hydrogen is considered green when it is produced from renewable energy sources, grid electricity that has a high average of renewable energy penetration or a low average of CO$_2$ intensity, or grid electricity when price is low. The ability to produce green hydrogen from grid electricity provides operation flexibility that can be exploited to co-minimise cost and CO$_2$ emissions. However, the operation flexibility raises the challenge of devising smart planning to achieve long-term hydrogen production targets within specific delivery periods.

Consequently, this research analysed a grid-connected renewable energy system considering benchmark and day-to-day operation performances. The benchmark performance considered a model with full foresight of various input data, whereas the day-to-day performance considered a model with two control modules, long-term and daily planners. The long-term planner adopted a novel approach which instead of long-term forecast utilised historical data short-term forecasts to determine hydrogen production for the upcoming day. The long-term planner also considered the remaining time and hydrogen that needs to be produced within a delivery period to ensure that the required total hydrogen production was achieved at the end of a delivery period. Based on the output of the long-term planner, the daily planner determined the optimal plan, which determined the hydrogen production at every hour of the upcoming day. At the beginning of a new day, the hydrogen production plant executed the production plan determined by the daily planner the day before. The execution required importing or exporting electricity to the grid depending on the availability of renewable energy.

The performance analysis considered daily, weekly, monthly, and yearly deliveries. For both benchmark and day-to-day operation performances, longer delivery periods provided more flexibility for co-optimising cost and CO$_2$ emissions. Significant reductions of CO$_2$ emissions could be achieved with relatively small changes of the levelised cost of hydrogen. Extended delivery periods are also associated with higher uncertainty in day-to-day operation; for yearly delivery, the cost increased instead of decreasing when the objective was tuned more towards minimising cost. When the objective was more towards minimising cost, the levelised costs considering benchmark performance were slightly underestimated in comparison to those of the day-to-day performance. When the objective was tuned to minimise emissions, the specific CO$_2$ emissions considering benchmark performance were significantly underestimated (up to 60\%), in comparison to those of the day-to-day performance.

For the purpose of green hydrogen production, grid electricity in Denmark could not be considered as renewable energy despite the large penetration from wind and solar. Consequently, approximately 30\% of the produced hydrogen would not be considered green hydrogen according to the latest rules set by the European Union~\cite{Directorate_General_for_Energy}. Nevertheless for yearly delivery targets, all the produced hydrogen was significantly less CO$_2$ intensive compared to existing alternative hydrogen production methods. The analysis of the rules in~\cite{Directorate_General_for_Energy} showed lack of transparency where in some cases hydrogen could be considered green despite being more CO$_2$ intensive than non-green hydrogen that were less CO$_2$ intensive. Hence, two main recommendations were proposed: first, the adoption of transparent accounting of hourly CO$_2$ emissions associated with hydrogen production, and second, adoption of a lower threshold for specific CO$_2$ emissions, the recommended threshold was 1.5~kg of CO$_2$ per kg of hydrogen produced.  

The main aim of this research was the development of the long-term planner that determined hydrogen production for the next day, hence, the results and recommendations should be interpreted cautiously considering that they are based on a single case study that considered specific system configuration and capacities, site location, hydrogen production target, and perfect short-term forecast. The results are also based on historical data that may not capture the current dynamics of the Danish electricity system, which is changing rapidly to achieve ambitious CO$_2$ reduction targets. Future research could analyse the results sensitivity to the assumptions adopted in this research. Future research could also consider improving the system performance by developing new approaches for the long-term planner.

In conclusion, this research presented a novel long-term planner that allows optimising the day-to-day operation of a grid-connected hybrid renewable energy plant for hydrogen production. The CO$_2$ emissions associated with hydrogen production can be significantly underestimated if day-to-day operation is ignored. With suitable control, significant reductions of CO$_2$ emissions with relatively small increase of the levelised cost of hydrogen can be achieved. Despite grid electricity not being considered as renewable energy for hydrogen production in Denmark, hydrogen production from both renewable energy and electricity grid is already more environmentally friendly than existing alternative hydrogen production methods. This research shows the importance of proper control of a hydrogen production plant in order to provide realistic day-to-day performance and to advance the transition towards sustainable energy systems, paving the way for a greener and resilient future.

\begin{acknowledgments}
Sleiman Farah received funding from Ørsted A/S and The Danish Energy Technology Development and Demonstration Program under grant number 64020-2120. Neeraj Bokde received funding from Apple Inc. as part of the APPLAUSE bio-energy collaboration with Aarhus University, Denmark. Gorm Bruun Andresen was the project responsible for the above grants. The authors acknowledge the contribution of Dr Tim T. Pedersen to the early stages of the research.

\end{acknowledgments}

\appendix

\section{Optimisation model details}
\label{A:Optimisation model details}

This section presents the optimisation model of the hydrogen production system. The model is developed in Python~\cite{van1995python}, and the co-optimisation of electricity cost and CO$_2$ emissions is resolved using Gurobi~\cite{gurobi} to determine the optimal power flows.
The optimisation model is a standard dispatch optimisation for the spot market. 

\begin{itemize}
   \item{DC bus power balance}
\end{itemize}
\begin{equation}
\label{E:DCBalance}
    g_{1,t} = g_{3dc,t} \le G_3/\eta_{i}
\end{equation}
\nomenclature{$g_{1,t}~(MW)$}{Solar power supplied to the DC bus}
\nomenclature{$g_{3,t}~(MW)$}{Power from the DC bus to the inverter}
\nomenclature{$G_3~(MW)$}{Power output capacity of the inverter}

where $g_{1,t}$, $g_{3dc,t}$, $G_5$, and $\eta_{i}$ represent the solar power supplied to the DC bus, inverter power (DC side), inverter capacity, and inverter efficiency respectively. The subscript $t$ represents hour of the year.

\begin{itemize}
    \item Solar power output considering curtailment
\end{itemize}
\begin{equation}
\label{E:SolarBalance}
    g_{1,t} = G_1 \times CF_{G_1,t} - g_{1c,t} \ge 0
\end{equation}
\nomenclature{$G_{1}~(MW)$}{Solar power capacity}
\nomenclature{$CF_{G_1,t}~$(-)}{Capacity factor of solar power at time t}
\nomenclature{$g_{1c,t}~(MW)$}{Curtailed solar power}

where $G_{1}$, $CF_{G_1,t}$, and $g_{1c,t}$ represent the solar capacity, solar capacity factor, and curtailed solar power respectively.
\begin{itemize}
    \item AC bus power balance 
\end{itemize}
\begin{equation}
    g_{2,t} + g_{3ac,t} - g_{4,t} + g_{5i,t} - g_{5e,t} = 0
\end{equation}
\nomenclature{$g_{2,t}~(MW)$}{Wind power supplied to the AC bus}
\nomenclature{$g_{6,t}~(MW)$}{Power from ($+$) or to ($-$) the AC bus}
\nomenclature{$g_{7,t}~(MW)$}{Power exported ($+$) or imported ($-$) from the grid}
\nomenclature{$G_6~(MW)$}{Power output capacity of the power converter (rectifier)}

where $g_{2,t}$, $g_{3ac,t}$, $g_{4,t}$, $g_{5i,t}$, and $g_{5e,t}$ represent the wind power supplied to the AC bus, inverter power output (AC side), electrolyser input power, grid power imported from the grid, and grid power exported to the grid respectively.

\begin{itemize}
    \item Wind power output considering curtailment
\end{itemize}
\begin{equation}
\label{E:WindBalance}
    g_{2,t} = G_2 \times CF_{G_2,t} - g_{2c,t} \ge 0
\end{equation}
\nomenclature{$G_{2}~(MW)$}{Wind power capacity}
\nomenclature{$CF_{G_2,t}~$(-)}{Capacity factor of solar power at time t}
\nomenclature{$g_{2c,t}~(MW)$}{Curtailed wind power}

where $G_{2}$, $CF_{G_2,t}~$, and $g_{2c,t}$ represent the wind capacity, wind capacity factor and curtailed wind power respectively.

\begin{itemize}
    \item Power limit of the grid connection
\end{itemize}

\begin{align}
    g_{5i,t} \le G_{5}\\
    g_{5e,t} \le G_{5}  
\end{align}
where $G_{5}$ is the grid connection capacity of power import and export to the grid.

\begin{itemize}
    \item Inverter power balance
\end{itemize}
\begin{equation}
    g_{3ac,t} = -g_{3dc,t} \times \eta_{i}
\end{equation}
\nomenclature{$\eta_{i}~$(-)}{Efficiency of converting DC to AC power}

\nomenclature{$\eta_{r}~$(-)}{Efficiency of converting AC to DC power}

\begin{itemize}
    \item Electrolyser energy balance
\end{itemize}
\begin{equation}
\label{E:g_{H_2,t}}
    g_{H_2,t} =  g_{4,t} \times \eta_{LHV_{H_2}}
\end{equation}
\nomenclature{$ECT_{min}$~(MWh)}{Minimum threshold of the thermo-mechanical energy storage content}
\nomenclature{$ECT_t$~(MWh)}{Energy content of StorageX}
\nomenclature{$\eta_{4,d}~$(\%)}{StorageX discharge efficiency}
\nomenclature{$L~$(-)}{Percentage of StorageX energy content lost to the environment each day}

where $g_{H_2,t}$ and $\eta_{LHV_{H_2}}$ represent the hydrogen power and the electrolyser efficiency based on the low heat value of hydrogen ($LHV_{H_2}$) respectively.

\begin{itemize}
    \item Electrolyser power limit
\end{itemize}
\begin{equation}
    g_{4,t} \le G_{4}
\end{equation}
where $G_{4}$ represents the electrolyser capacity.
\begin{itemize}
    \item Electrolyser ramping limits 
\end{itemize}
\begin{align}
    g_{4,t+1} - g_{4,t} &\le G_{4}\times rru\\
    g_{4,t+1} - g_{4,t} &\le G_{4}\times rruc\\
    g_{4,t} - g_{4,t+1} &\le G_{4}\times rrd
\end{align}
where $rru$, $rruc$, and $rrd$ are the ramp up rate from warm state, ramp up rate from cold state, and ramp down rate respectively.
\begin{itemize}
    \item Hydrogen mass output
\end{itemize}
\begin{equation}
        m_{H_2,t} =  \frac{3600 \times g_{H_2,t} \times \Delta t}{LHV_{H_2}}
    \end{equation}

where $m_{H_2,t}$ is the mass of hydrogen. The mass of hydrogen produced during a period ($T$) should satisfy a minimum demand ($m^T_{H_2}$) for that period as expressed in  Eq.~\ref{E:constraint_annual_mass_production_h2}
\begin{equation}
    \sum_{t=1}^{T/\Delta t} m_{H_2,t} \ge m^T_{H_2}
    \label{E:constraint_annual_mass_production_h2}
\end{equation}

\nomenclature{$\Delta t~(h)$}		{Time step}
\nomenclature{$t$~(-)}{Discrete time variable}
\nomenclature{$CF_{G_7}$~(-)}{Capacity factor of the grid connection}

The constraint in Eq.~\ref{E:constraint_annual_mass_production_h2} is to control the hydrogen production for shorter delivery periods.

\begin{itemize}
    \item System cost
\end{itemize}
The system cost ($\hat{C}_s$) consists of the capital cost ($\hat{C}_c$) and operation cost ($\hat{C}_o$) as shown in Eq.~\ref{E:system_cost}
\begin{equation}
\label{E:system_cost}
    \hat{C}_s = \hat{C}_c + \hat{C}_o
\end{equation}

The capital cost is calculated as per Eq.~\ref{E:captial_cost}
\begin{equation}
    \hat{C}_c = \sum_{k=1}^{n} \hat c^a_k\times G_k
    \label{E:captial_cost}
\end{equation}

where $n$, $\hat c^a_k$ and $G_k$ are the number of technologies in the system, the annualised cost per unit capacity of technology $k$ and the capacity of technology $k$ respectively. The $\hat c^a_k$ is calculated as a function of the annual fixed cost per unit capacity ($\hat c_{f,k}$), the investment cost per unit capacity ($\hat c_{i,k}$), the discount rate ($d$), and the technology lifetime ($L_k$) as shown in Eq.~\ref{E:AC}
\begin{equation}
    \hat c^a_k = \hat c_{f,k} + \frac{\hat c_{i,k} \times d}{1-\left ( 1+d\right)^{-L_k}}
    \label{E:AC}
\end{equation}

The operation cost is calculated as per Eq.~\ref{E:operation_cost}
\begin{equation}
    \hat{C}_o = \sum_{k=1}^{n} \sum_{t=1}^{T/\Delta t} \hat c_{o,k}\times g_{k,t} \times \Delta t 
    \label{E:operation_cost}
\end{equation}
where $\hat c_{o,k}$, and $g_{k,t}$ are the operation cost per unit of operation power and operation power of technology $k$ respectively.

\begin{itemize}
    \item CO$_2$ cost
\end{itemize}
\begin{equation}
    \hat{C}_{CO_2} = \hat c_{CO_2} \times \Delta t \times \sum_{t=1}^{T/\Delta t}{g_{5i,t}\times I_{CO_2, t}}
    \label{E:co2_cost}
\end{equation}

where $\hat c_{CO_2}$, $g_{5i,t}$, and $I_{CO_2, t}$ represent the cost of emitting one kg of CO$_2$, the power imported from the grid, and the CO$_2$ intensity in the grid (kg/MWh) respectively. This calculation assumes no financial revenue associated with offsetting CO$_2$ emissions through the export of electricity to the grid. This calculation of the CO$_2$ cost is motivated by the idea that exporting electricity to the grid does not rectify irreversible environmental damages caused by importing non-renewable electricity from the grid.  

\begin{itemize}
    \item Electricity cost
\end{itemize}
Unlike the CO$_2$ cost, the electricity cost includes the revenue from exporting electricity to the grid as shown in Eq.~\ref{E:electricity_cost}
\begin{equation}
\label{E:electricity_cost}
    \hat C_e = \Delta t\times \sum_{t=1}^{T/\Delta t}{(g_{5i,t} - g_{5e,t}) \times \hat c_{e,t}}
\end{equation}
where $g_{5i,t}$ and $g_{5e,t}$ represent the electrical power imported and exported to the grid respectively, and $\hat c_{e,t}$ represents the electricity cost in the spot market.

\section{Electrolyser efficiency}
\label{A:electrolyser_efficiency}
\begin{figure}
	\centering{\includegraphics[width=0.5\textwidth]{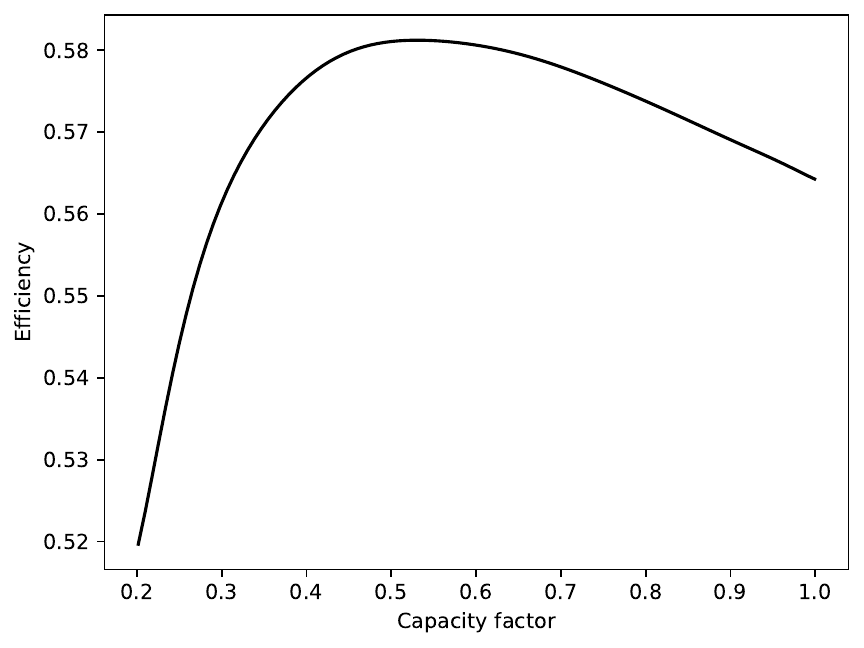}}
	\caption{Electrolyser efficiency ($\eta_{LHV_{H_2}}$) for different capacity factors ($f_{4,t} = g_{4,t}/G_4)$ \cite{liponi2022techno}.}
 \label{F:electrolyser_efficiency}
\end{figure}

The electrolyser efficiency varies non-linearly from approximately 0.52 to 0.565 as the capacity factor increases from 0.2 to 1.0. The highest efficiency is approximately 0.58, which corresponds to a capacity factor slightly larger than 0.5, as shown in Fig.~\ref{F:electrolyser_efficiency}. This efficiency curve suggests that a variable efficiency should be considered in the system model.

\begin{figure}
\centering{\includegraphics[width=0.5\textwidth]{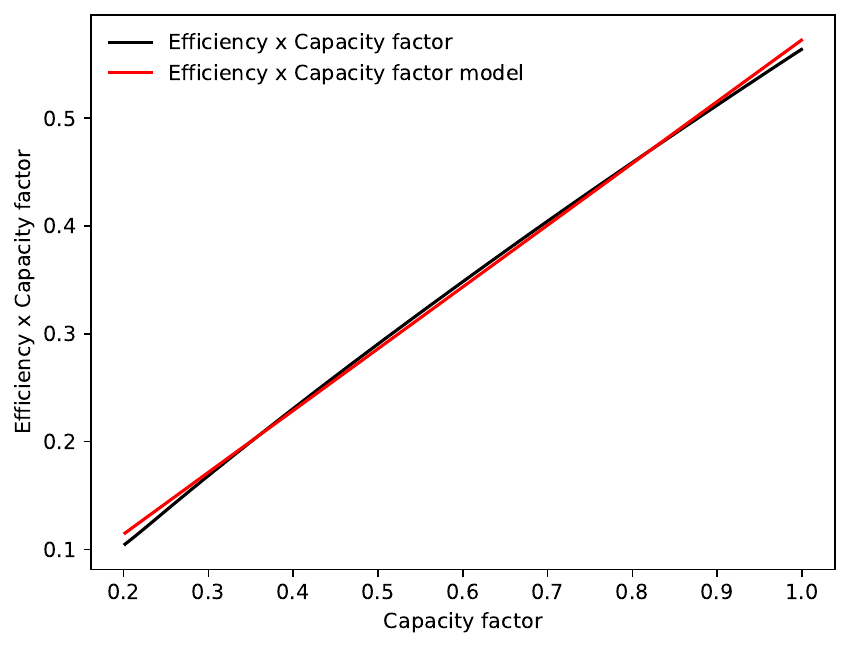}}
	\caption{Product of electrolyser efficiency and capacity factor ($f_{4,t} \times \eta_{LHV_{H_2}, t}$) for different capacity factors ($f_{4,t}$).}
 \label{F:electrolyser_product}
\end{figure}

The variable electrolyser efficiency introduces non-linearity in Eq.~\ref{E:g_{H_2,t}} as the efficiency becomes a function of the capacity factor, which is the ratio of the electrolyser power and power capacity. Therefore, to maintain the linearity of the system model, instead of modelling the efficiency and utilising Eq.~\ref{E:g_{H_2,t}} to calculate $g_{H_2,t}$, $g_{H_2,t}$ is considered as an optimisation decision variable that should still satisfy the constraint set by Eq.~\ref{E:g_{H_2,t}}. This constraint can be represented as shown in Eq.~\ref{E1:g_{H_2,t}}

\begin{equation}
    \label{E1:g_{H_2,t}}
    g_{H_2,t} = G_{4} \times f_{4,t} \times \eta_{LHV_{H_2}, t}
\end{equation}

where $f_{4,t}$ is the electrolyser capacity factor. The product $f_{4,t} \times \eta_{LHV_{H_2}, t}$ can be calculated based on the efficiency function in Fig.~\ref{F:electrolyser_efficiency}. The product $f_{4,t} \times \eta_{LHV_{H_2}, t}$ is shown in Fig.~\ref{F:electrolyser_product} as a function of $f_{4,t}$.

Fig.~\ref{F:electrolyser_product} reveals that $f_{4,t} \times \eta_{LHV_{H_2}, t}$ is practically a linear function of {$f_{4,t}$}; the function can be determined using regression analysis. The results of regression analysis shows that the slope and intercept of the linear function are 0.5734 and -0.0005 respectively. The p-value of the slope is less than 0.001 whereas the p-value of the intercept 0.685, which is higher than 0.05. These results indicate that the intercept can be omitted from the linear function. With the omission of the intercept, the adjusted slope value is 0.5726, and $f_{4,t} \times \eta_{LHV_{H_2}, t}$ can be presented as shown in Eq.~\ref{E2:g_{H_2,t}}  
\begin{equation}
    \label{E2:g_{H_2,t}}
    f_{4,t} \times \eta_{LHV_{H_2}, t} = 0.5726 \times f_{4,t}
\end{equation}

The result in Eq.~\ref{E2:g_{H_2,t}} suggests that the value of $\eta_{LHV_{H_2}, t}$ can be considered a constant equal to 0.5726 for capacity factors in the range of 0.2--1.0. The adopted value is $\eta_{LHV_{H_2}, t}=0.6$ across the capacity factors range of 0.0--1.0 to simplify the modelling. 

\section{Additional results}
\label{A:additional_results}
\begin{figure}
  % \centering
  \begin{subfigure}[t]{0.49\textwidth}
    % \centering
    \includegraphics[width=\textwidth]{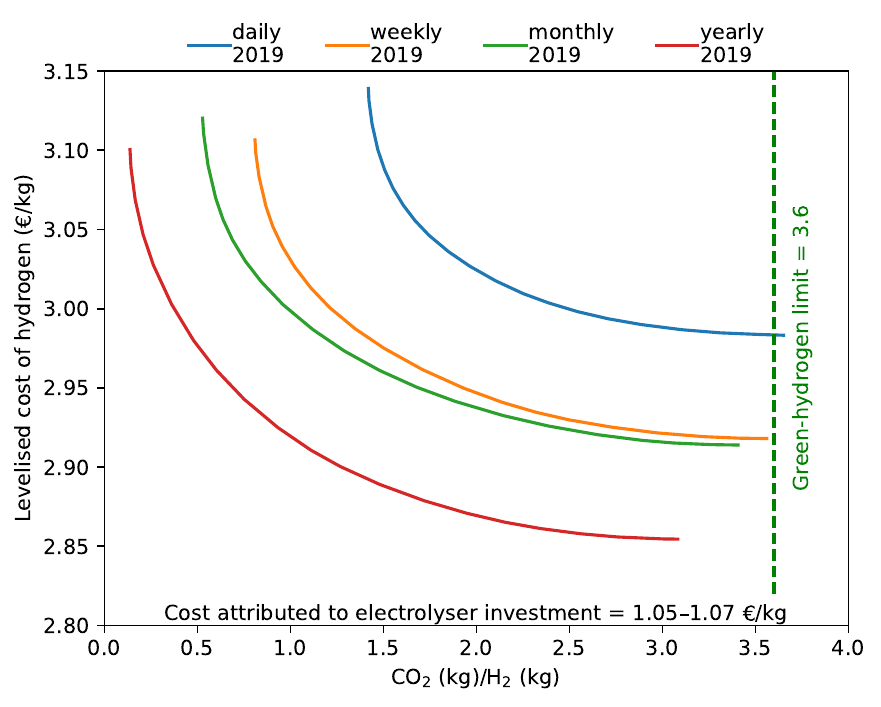}
    \caption{Benchmark}
    \label{F:benchmark_2019}
  \end{subfigure}
    \hfill
  \begin{subfigure}[t]{0.49\textwidth}
    % \centering
    \includegraphics[width=\textwidth]{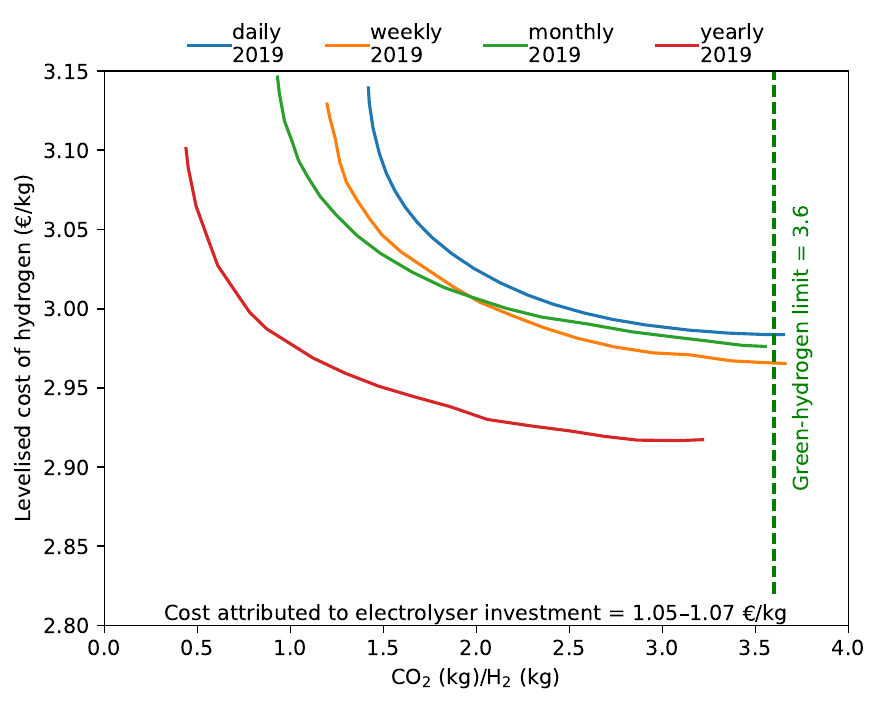}
    \caption{Day-to-day}
    \label{F:day-to-day_2019}
  \end{subfigure}
  \caption{Impact of specific CO$_2$ emissions and delivery frequencies on the levelised cost H$_2$ (year 2019).\hspace{\textwidth}}
  \label{F:Impact of specific CO$_2$ emissions and delivery frequency on the levelised cost H$_2$ (year 2019)}
\end{figure}

\begin{figure}
  % \centering
  \begin{subfigure}[t]{0.49\textwidth}
    % \centering
    \includegraphics[width=\textwidth]{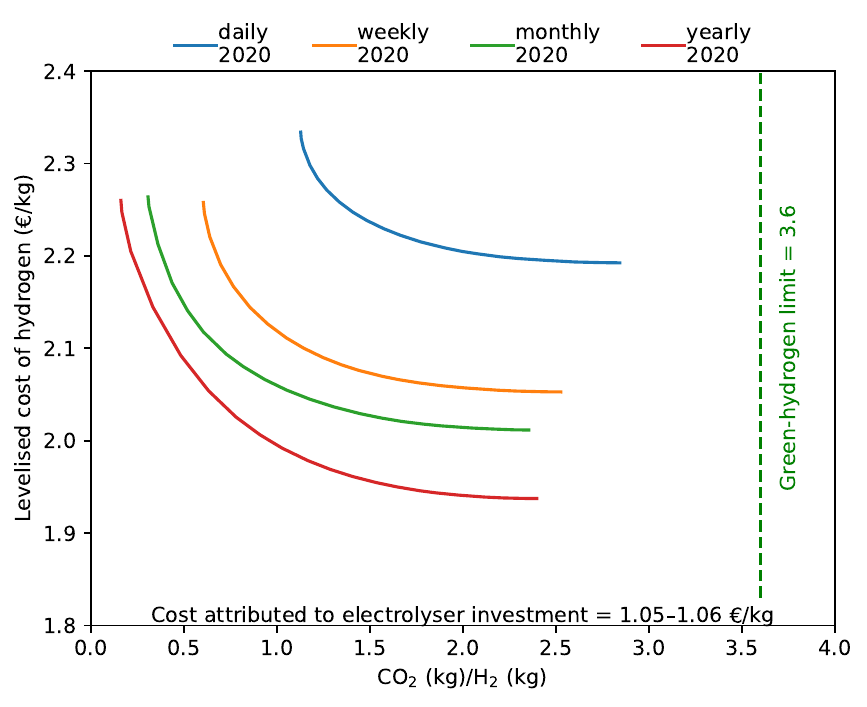}
    \caption{Benchmark}
    \label{F:benchmark_2020}
  \end{subfigure}
    \hfill
  \begin{subfigure}[t]{0.49\textwidth}
    % \centering
    \includegraphics[width=\textwidth]{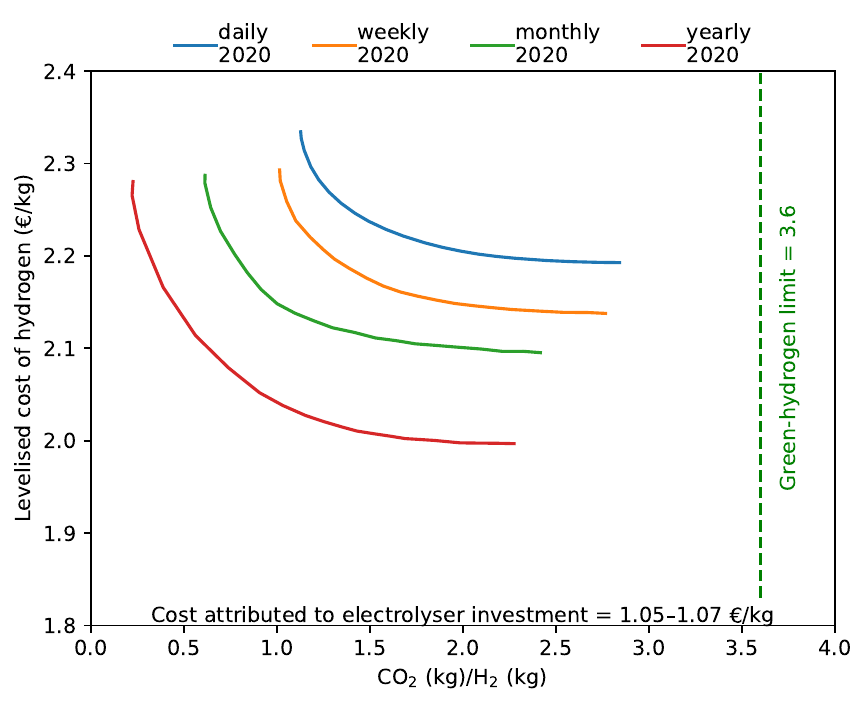}
    \caption{Day-to-day}
    \label{F:day-to-day_2020}
  \end{subfigure}
  \caption{Impact of specific CO$_2$ emissions and delivery frequencies on the levelised cost H$_2$ (year 2020).\hspace{\textwidth}}
  \label{F:Impact of specific CO$_2$ emissions and delivery frequency on the levelised cost H$_2$ (year 2020)}
\end{figure}

\begin{figure}
  % \centering
  \begin{subfigure}[t]{0.49\textwidth}
    % \centering
    \includegraphics[width=\textwidth]{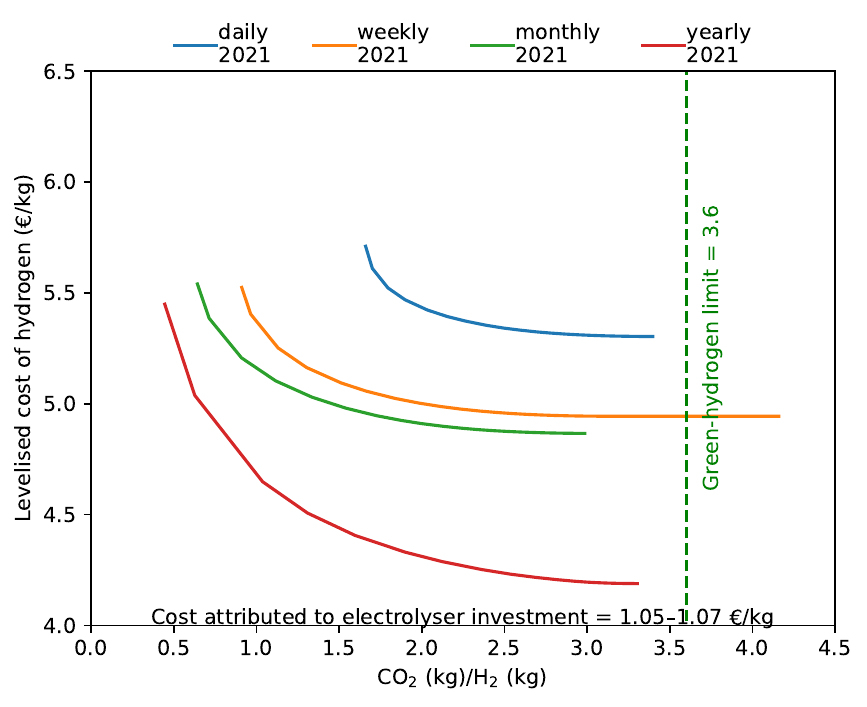}
    \caption{Benchmark}
    \label{F:benchmark_2021}
  \end{subfigure}
    \hfill
  \begin{subfigure}[t]{0.49\textwidth}
    % \centering
    \includegraphics[width=\textwidth]{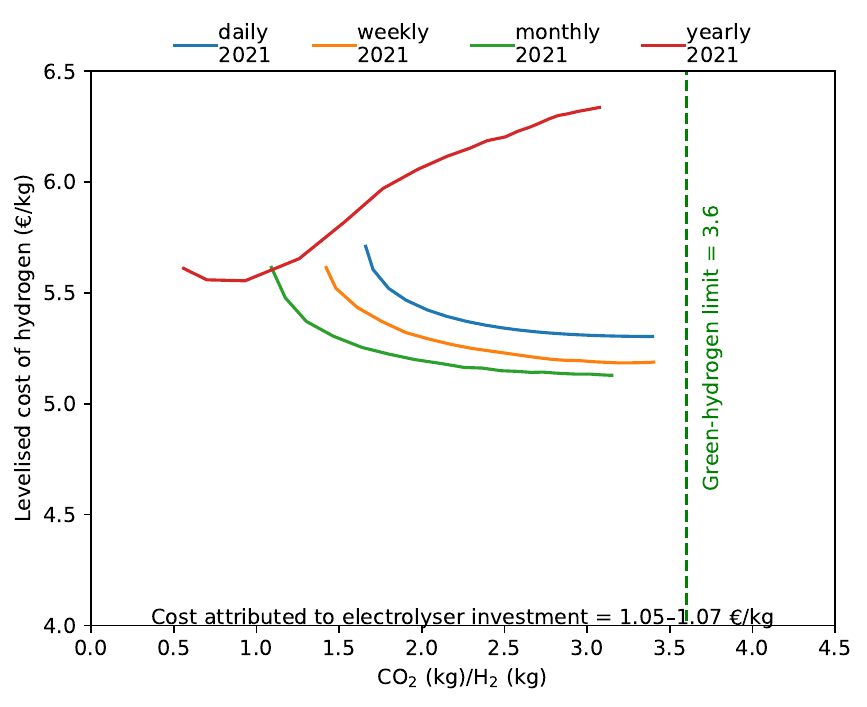}
    \caption{Day-to-day}
    \label{F:day-to-day_2021}
  \end{subfigure}
  \caption{Impact of specific CO$_2$ emissions and delivery frequencies on the levelised cost H$_2$ (year 2021).\hspace{\textwidth}}
  \label{F:Impact of specific CO$_2$ emissions and delivery frequency on the levelised cost H$_2$ (year 2021)}
\end{figure}

Benchmark and day-to-day performances considering weather conditions for years 2019, 2020, and 2021 are presented in Figs.~\ref{F:Impact of specific CO$_2$ emissions and delivery frequency on the levelised cost H$_2$ (year 2019)},~\ref{F:Impact of specific CO$_2$ emissions and delivery frequency on the levelised cost H$_2$ (year 2020)}, and~\ref{F:Impact of specific CO$_2$ emissions and delivery frequency on the levelised cost H$_2$ (year 2021)} respectively.

\section{Analysis of day-to-day performance for yearly delivery}
\label{A:analysis_of_day_to_day_performance_for_yearly_delivery}

\begin{figure}
    \centering
    \begin{subfigure}[b]{0.49\textwidth}
        \centering
        \includegraphics[width=\textwidth]{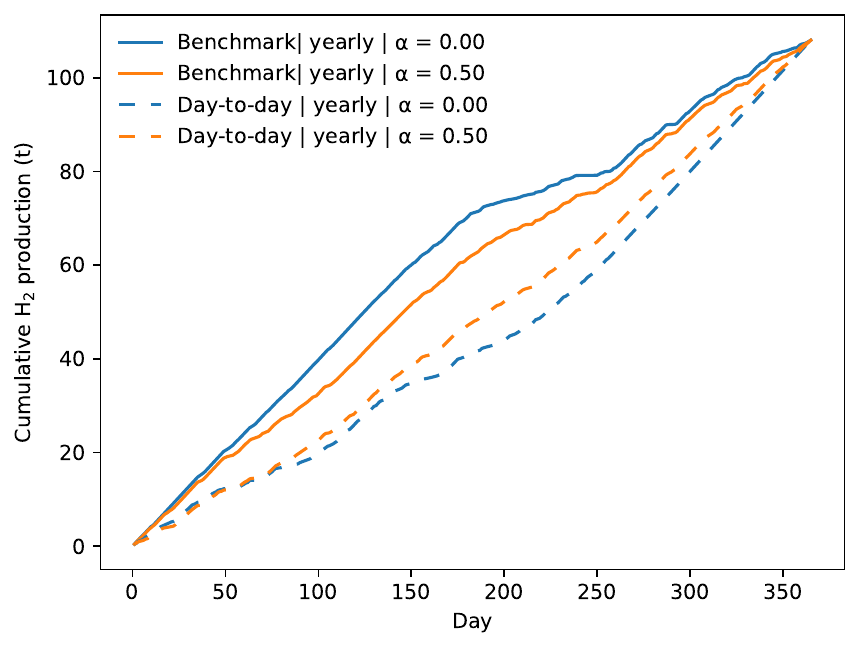}
        \caption{Comparison of cumulative H$_2$ produced (year 2018, yearly delivery, shrinking-window long-term planner)}
        \label{F:Comparison of cumulative H$_2$ produced}
    \end{subfigure}
    \hfill
    \begin{subfigure}[b]{0.49\textwidth}
        \centering
        \includegraphics[width=\textwidth]{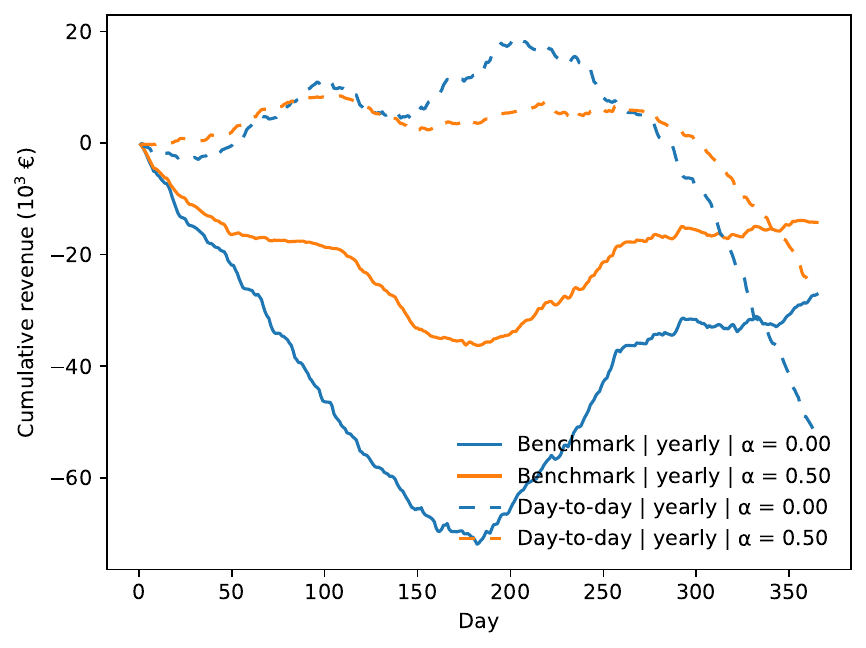}
        \caption{Comparison of cumulative net revenue (year 2018, yearly delivery, shrinking-window long-term planner)}
        \label{F:Comparison of cumulative net revenue}
    \end{subfigure}
    \hfill
    % \begin{subfigure}[b]{0.45\textwidth}
    %     \centering
    %     \includegraphics[width=\textwidth]{co2_emission_2018_yearly_0.0_1.0_benchmark_and_shrinking_window.pdf}
    %     \caption{Comparison of cumulative CO$_2$ emissions (year 2018, yearly delivery, shrinking-window long-term planner)}
    %     \label{F:Comparison of cumulative CO$_2$ emissions}
    % \end{subfigure}
    \caption{Comparison of different metrics.}
    \label{fig:comparison}
\end{figure}

To better understand the result of yearly delivery performance, the cumulative hydrogen production and net revenue are analysed considering the CO$_2$ weighting factors that correspond to the lowest levelised cost and the highest specific CO$_2$ emissions. Considering the day-to-day operation in Fig.~\ref{F:Comparison of cumulative H$_2$ produced}, for most of the year, hydrogen production with the aim of minimising the cost ($\alpha=0$) is delayed, hence, the cumulative hydrogen production is lower than that for co-minimising cost and CO$_2$ emissions ($\alpha=0.5$). The hydrogen production delay does achieve a higher net revenue up to day 250 (Fig.~\ref{F:Comparison of cumulative net revenue}). However, towards day 200 onwards, the system operates almost continuously to be able to achieve the required hydrogen target by the end of the delivery period. The continuous operation is reflected by the linear trend of the cumulative hydrogen produced. The continuous operation eliminates the ability of benefiting from variations of electricity prices, resulting in a sharp decline of the cumulative revenue towards the end of the year.

In addition, the cumulative hydrogen production shows a significant difference between the benchmark performance and the day-to-day performance especially when the objective is cost minimisation only (blue lines in Fig.~\ref{F:Comparison of cumulative H$_2$ produced}). The benchmark performance shows a higher hydrogen production rate in the first half of the year in comparison to that of the second half of the year. This production profile is in contrast with the production profile of the day-to-day performance. When the objective is cost and CO$_2$ emissions co-minimisation only (orange lines in Fig.~\ref{F:Comparison of cumulative H$_2$ produced}), the difference between the benchmark and the day-to-day performances is smaller. Similarly, considering the cumulative revenue results ($\alpha=0$), benchmark performance shows a rapid decline of the cumulative net revenue in the first half of the year. However, part of the revenue is recovered during the second half of the year. This cumulative net revenue pattern is significantly different from that of the day-to-day operation where a sharp decline occurs in the second half of the year. A relatively smaller difference can be observed between the benchmark and the day-to-day performances when the aim to co-minimise cost and CO$_2$ emissions ($\alpha=0.5$).

% \nocite{*}

\bibliography{Bibliography}

\end{document}